%% file: cameraready.tex
\documentclass{article}

\usepackage{microtype}
\usepackage{graphicx}
\usepackage{subfig}
\usepackage{booktabs} 

\usepackage{hyperref}

\usepackage[accepted]{icml2024}

\usepackage{amsmath}
\usepackage{amssymb}
\usepackage{mathtools}
\usepackage{amsthm}

\usepackage[capitalize,noabbrev]{cleveref}

\theoremstyle{plain}

\theoremstyle{definition}

\theoremstyle{remark}

\usepackage{booktabs} 
\usepackage{siunitx} 
\usepackage[textsize=tiny]{todonotes}
\input{commands}

\icmltitlerunning{\dpolar codes}

\begin{document}

\twocolumn[
\icmltitle{\dpolar: Inventing Nonlinear Large-Kernel Polar Codes via Deep Learning}




\begin{icmlauthorlist}
\icmlauthor{S Ashwin Hebbar*}{princeton}
\icmlauthor{Sravan Kumar Ankireddy*}{texas}
\icmlauthor{Hyeji Kim}{texas}
\icmlauthor{Sewoong Oh}{wash}
\icmlauthor{Pramod Viswanath}{princeton}
\end{icmlauthorlist}

\icmlaffiliation{princeton}{Princeton University}
\icmlaffiliation{wash}{University of Washington}
\icmlaffiliation{texas}{University of Texas at Austin}

\icmlcorrespondingauthor{Ashwin Hebbar}{hebbar@princeton.edu}

\icmlkeywords{Machine Learning, ICML}

\vskip 0.3in
]



\printAffiliationsAndNotice{*Lead authors}  
\begin{abstract}


Progress in designing channel codes has been driven by human ingenuity and, fittingly, has been sporadic. Polar codes, developed on the foundation of Arikan's polarization kernel, represent the latest breakthrough in coding theory and have emerged as the state-of-the-art error-correction code for short-to-medium block length regimes. In an effort to automate the invention of good channel codes, especially in this regime, we explore a novel, non-linear generalization of Polar codes, which we call \dpolar codes. \dpolar codes extend the conventional Polar coding framework by utilizing a larger kernel size and parameterizing these kernels and matched decoders through neural networks. Our results demonstrate that these data-driven codes effectively leverage the benefits of a larger kernel size, resulting in enhanced reliability when compared to both existing neural codes and conventional Polar codes.
Source code is available at \href{https://www.github.com/hebbarashwin/deeppolar}{ this link}.

%
%
%

\end{abstract}

\section{Introduction}
\label{introduction}






Reliable digital communication is a primary workhorse of the information age. To ensure reliable communication over a noisy channel, it is common to introduce redundancy in the transmitted data to enable faithful reconstruction of the message by receivers. This crucial process, known as error correction coding (channel coding), lies at the heart of both wired (Ethernet, cable) and wireless (cellular, WiFi, satellite) communication systems. Over the past seven decades, a significant research thrust has focused on designing reliable codes (consisting of an encoder-decoder pair) that achieve good reliability whilst having an efficient decoder. The canonical setting is one of point-to-point reliable communication over the additive white Gaussian noise (AWGN) channel, and the performance of a code in this setting is its gold standard. The figure of merit can be precisely measured: bit error rate (BER) measures the fraction of input bits that were incorrectly decoded; block
error rate (BLER) measures the fraction of times at least
one of the original data bits was incorrectly decoded. 

The field of coding theory has experienced sporadic yet significant breakthroughs, largely propelled by human ingenuity. Polar codes, invented by Arikan in 2009 \cite{arikan2009channel}, is one of the most profound developments in coding theory, significantly revitalizing the field. Polar codes, a combination of algebraic and graphical coding structures, are the first class of codes with a deterministic construction proven to achieve Shannon capacity. Importantly, this is achieved with low-complexity encoding and decoding. The impact of Polar codes is evident from their integration into the 5G standards within just a decade of their proposal – a remarkably swift timeline considering that it typically takes several decades for new coding methods to be incorporated into cellular standards ~\cite{3GPP38212, bioglio2020design}. 

The basic building block of Polar codes is a binary matrix $G = \begin{bmatrix} 1 & 0 \\ 1 & 1 \end{bmatrix}$, called the polarization \textit{kernel}. The generative matrix, which characterizes a linear code, is obtained by several Kronecker products of $G$ with itself. This construction of polar codes gives rise to a remarkable phenomenon known as ``channel polarization".  This process transforms $n$ views of a binary memoryless channel into $n$ synthetic ``bit channels", each with distinct reliability. As the block length $n$ grows asymptotically large, these bit channels become \textit{polarized}, becoming either completely noiseless or completely noisy. Polar encoding proceeds by sending information bits in the noiseless bit channels while the noisy input bits are ``frozen" to a known value. Capacity is achieved via the sequential successive cancellation (SC) decoder.

Polar codes and SC decoding are optimal for large asymptotic blocklengths; however, practical finite-length performance is lacking. Recognizing this limitation, recent research has focused on augmenting the polar encoder and improving the decoding performance. Specifically, the concatenation of cyclic redundancy check (CRC) with polar codes improves distance properties. Combining this with successive cancellation list (SCL) decoding markedly enhances decoding performance \cite{tal2015list, niu2012crc}. Consequently, Polar codes have emerged as the state-of-the-art in the short-to-medium block length regime, leading to their inclusion in 5G standards. Nonetheless, SCL decoding introduces considerable decoding complexity and latency, marking a trade-off between performance and computational efficiency. 

In parallel, there have been efforts to improve polar encoders. One approach involves modifying the Polar encoding structure. A noteworthy improvement is polarization-adjusted convolutional (PAC) codes \cite{arikan2019sequential}, which use convolutional precoding before polar coding. Remarkably, PAC codes approach finite-length information-theoretic bounds for binary codes in the short block length regime. However, the practical application of PAC codes is limited by their high decoding complexity, necessitating extremely large list sizes for effective decoding. 

An alternative idea involves increasing the size of the polarization kernel. Indeed, \cite{korada2010polar} prove that polarization holds for all kernels provided they are not unitary and not upper triangular under any column permutation. Further, one can find large polarization kernels ($\ell \ge 8$) that achieve faster polarization (characterized by the "scaling exponent"). 
\textcolor{black}{Despite these theoretical advantages, Polar codes with large kernels are not preferred in practice due to their increased decoding complexity. Nevertheless, there is recent work addressing these challenges \cite{trifonov2023design} based on specially designed polarization kernels.}

\textcolor{black}{In this work, we aim to address the question: How can we automate the search for good codes? Indeed, it is possible by parameterizing and learning both the encoder and decoder using neural networks. 
However, constructing effective non-linear codes using this approach is highly challenging: it is well-documented in literature that naively parameterizing with off-the-shelf neural architectures often results in performance worse than even repetition codes, as elaborated further 
in \prettyref{app:fcnn}. Rather, a more promising approach is to design neural architectures that enable structured redundancy. Specifically, our work delves into the innovative intersection between algebraic coding theory and machine learning by exploring non-linear generalizations of polarization-driven structures.} This can be achieved by parameterizing each kernel by a neural network, combining the information-theoretic properties of polarization-driven code structures with the adaptability and learning capabilities of deep learning. This interplay between algebraic coding structures and deep learning is a relatively unchartered territory. Our algorithm builds upon the groundwork laid by \cite{Makkuva2021}, who introduced KO codes as a non-linear generalization of RM codes. Although this work marked a significant leap forward, it is limited by its dependence on Reed-Muller (RM) encoding-decoding schemes, which restricted its applicability across a broader range of rates. Importantly, the architecture of KO codes proved inapplicable for Polar codes, being only scalable upto a (64,7) code. In contrast, our work proposes \dpolar codes, a non-linear generalization of the encoding and decoding structures of Polar codes (which encompass RM codes as a special case). This allows us to scale seamlessly to various rates and block lengths.

\begin{figure}[ht]
\begin{center}
\includegraphics[width=\columnwidth]{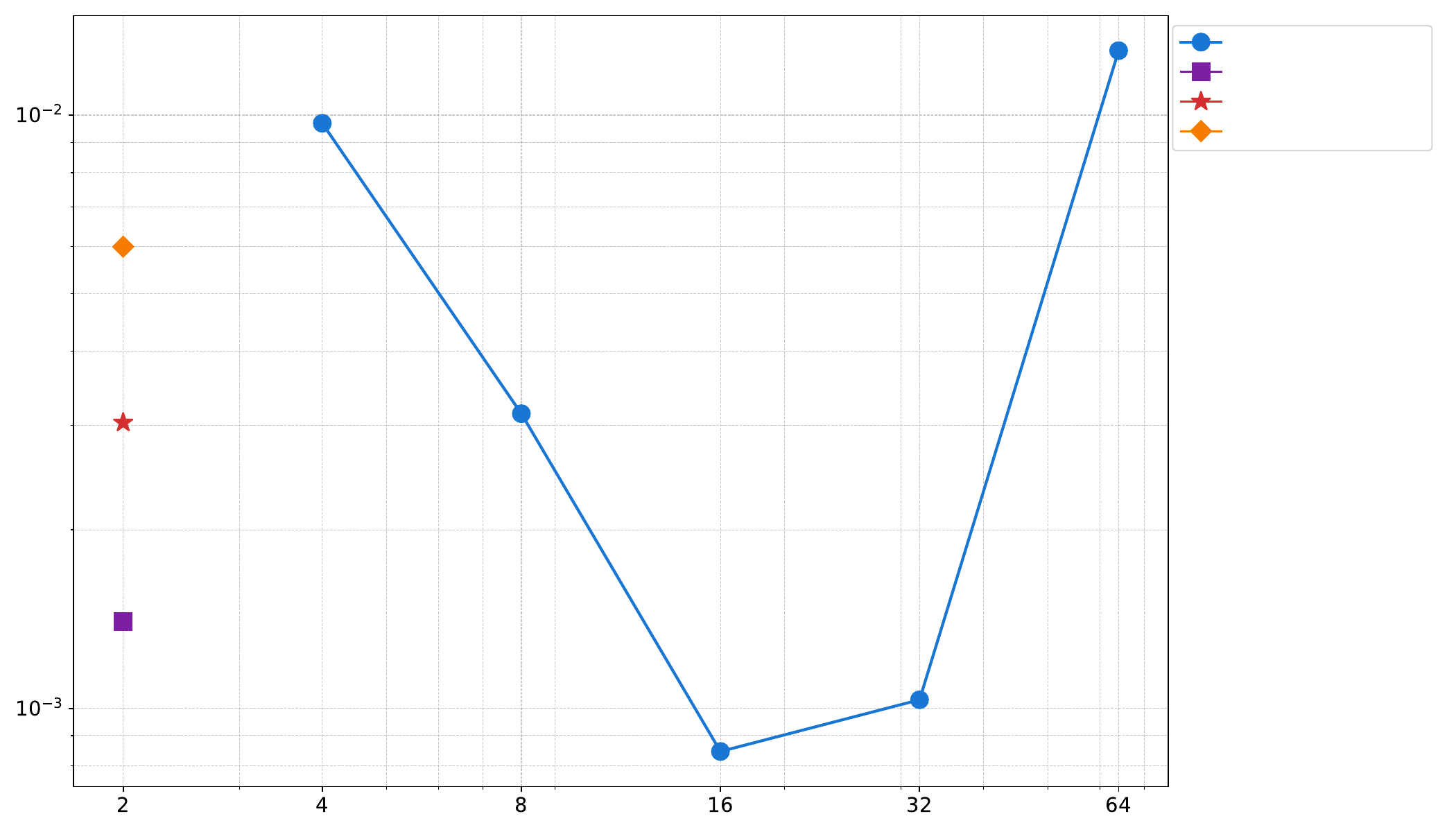}
\put(-37,126){\fontsize{4}{5}\selectfont \dpolar(256,37)}
\put(-37,121){\fontsize{4}{5}\selectfont KO(8,2)}
\put(-37,116){\fontsize{4}{5}\selectfont $\polar$(256,37)}
\put(-37,112){\fontsize{4}{5}\selectfont RM(8,2}
\put(-160,-5){\footnotesize Kernel size $\ell$}
\put(-240,70){\rotatebox[origin=t]{90}{\footnotesize Bit Error Rate}}
\caption{\dpolar ($n$=256, $k$=37) with appropriate kernel sizes (e.g., $\ell=16,32$) outperforms classical Reed-Muller, Polar, and state-of-the-art neural KO codes \cite{Makkuva2021} on AWGN channels with -2dB SNR}
\label{fig:kernel_size}
\end{center}
\vspace{-1em}
\end{figure}
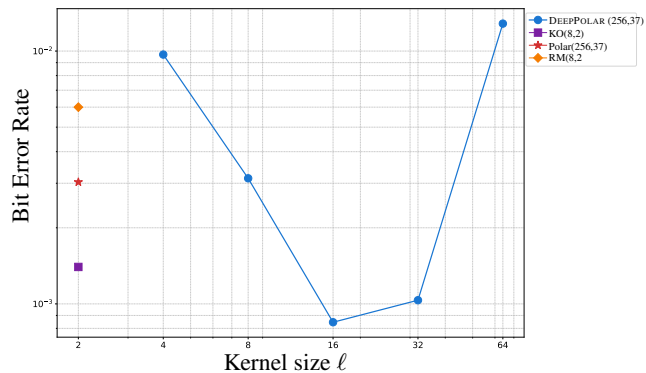

The core innovation of \dpolar lies in its {\em utilization of larger-size kernels}, which enables the neural network to explore an expansive function space. This is complemented by a matching SC-like neural decoder, which is jointly trained on samples drawn from an AWGN channel. Larger kernel sizes are pivotal for \dpolar achieving substantial performance improvements on the canonical AWGN channel, as evidenced in \prettyref{fig:kernel_size} for the case of $n=256, k=37$. Through extensive empirical studies, we find that a kernel size $\ell = \sqrt{n}$ is most effective in balancing bias and variance, enabling it to achieve significantly lower bit error rates over the baseline Polar and RM codes, as well as the KO($8,2$) code, SOTA at this block length and rate. Additionally, we design a training curriculum that builds upon the inherent nested hierarchy of polar coding structures to accelerate training.  
This combination of principled coding-theoretic structures with a targeted training methodology is designed to enable effective generalization across the vast space of messages. In summary, we make the following contributions:
\looseness=-1
\begin{itemize}
    \item We propose \dpolar, a novel generalization of Polar codes via large non-linear NN-based kernels.
    \item \dpolar outperforms the classical Polar and Reed-Muller codes, state-of-the-art binary linear codes, as well as KO codes, state-of-the-art ML-based codes, whilst scaling to various code rates.
    \item We develop a principled curriculum-based training methodology that allows \dpolar to generalize to the challenging high SNR regime.
\end{itemize}

\section{Problem formulation and Background}

\subsection{Channel coding}
Channel coding is a technique to add redundancy to the transmission to make it robust against noise added by the communication channel. More precisely, let $\bu =(u_0,\ldots, u_{k-1}) \in \{0,1\}^k$ denote a block of \emph{information/message} bits that we wish to transmit. A code consists of an encoder and decoder pair. The encoder $g_\phi:\binary^k \to \binary^n$ maps these message bits into a binary codeword $\bx$ of length $n$, \ie $\bx=g(\bu)$. The codewords are mapped to real/complex values by modulation (eg, Binary Phase Shift Keying (BPSK)). The channel, denoted as $W_{Y|X}(\cdot|\cdot)$, corrupts the codeword $\bx$ to its noisy version $\by \in \reals^n$. Upon receiving the corrupted codeword, the decoder $f_\theta$ estimates the message bits as $\hat{\bu}=f_\theta(\by)$. The performance of the code is measured using standard error metrics such as Bit-Error-Rate (BER) or Block-Error-Rate (BLER): $\mathrm{BER}(g_\phi, f_\theta) \define (1/k) \sum_i \pprob{\hat{u}_i \neq u_i}$, whereas $\mathrm{BLER}(g_\phi, f_\theta) \define \pprob{\hat{\bu} \neq \bu}$.

\begin{figure}[ht]
\begin{center}
\includegraphics[width=\columnwidth, trim={0 12cm 0 8cm},clip]{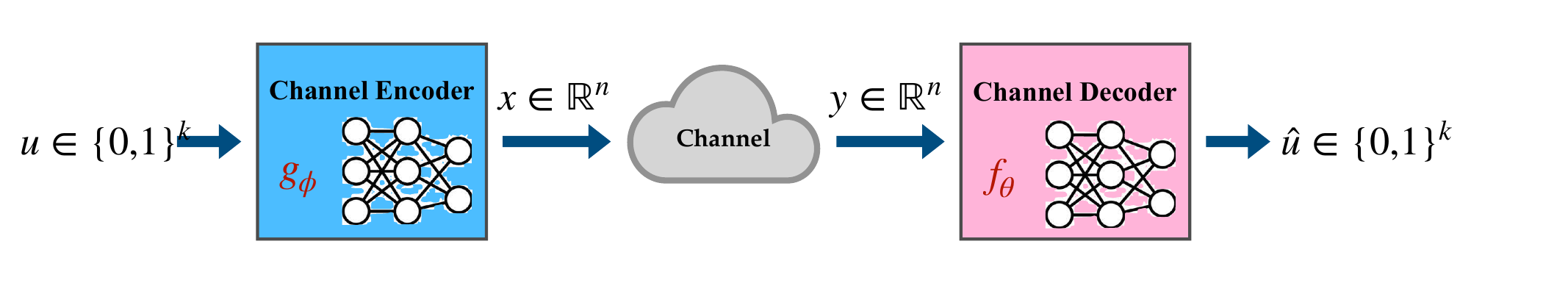}
\caption{Channel coding via deep learning}
\label{fig:ber_256_37}
\end{center}
\vskip -0.2in
\end{figure}

\subsection{Polar codes}

\begin{figure*}[ht]
\begin{center}
\includegraphics[width=0.8\linewidth, trim={1cm 8cm 16cm 6cm},clip]{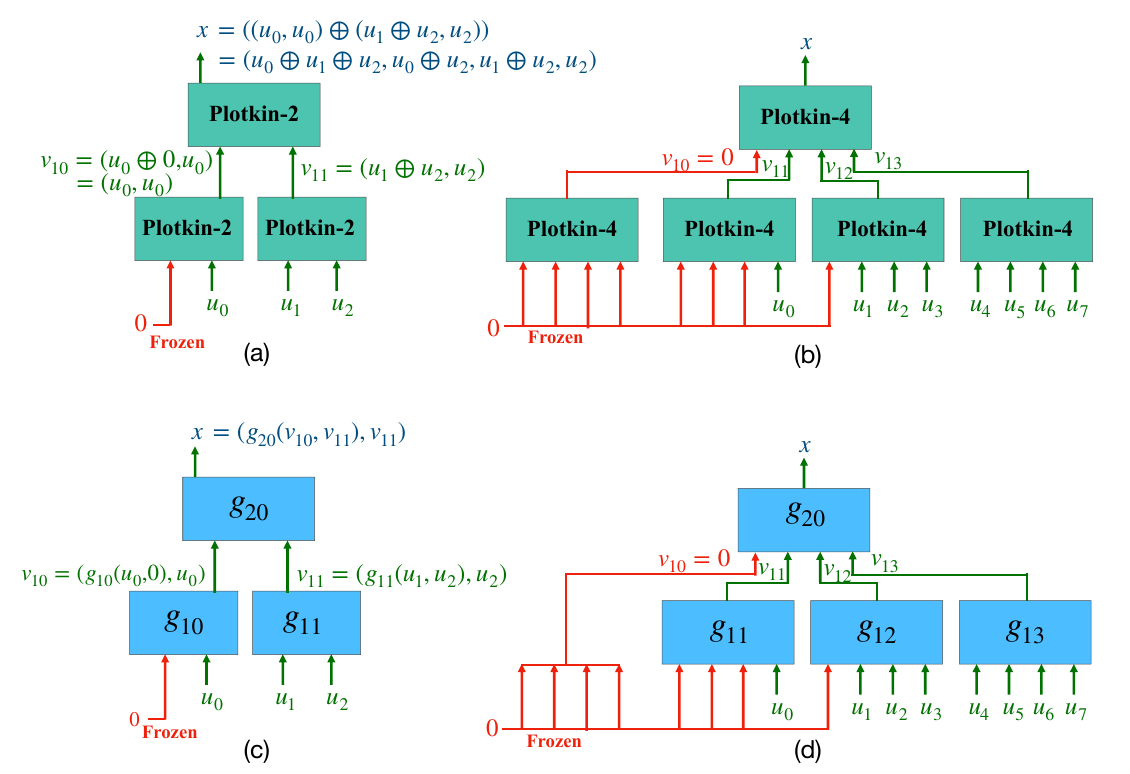}
\caption{(a) Polar$(4,3)$ encoding structure using the standard $2 \times 2$ kernel. Encoding is performed recursively on the Plotkin tree. (b) Polar$(16,8)$ encoding using $4\times 4$ kernels. (c) \dpolar($4,3,\ell=2)$ replaces the xor operation in Plotkin - $2 \times 2$ by neural networks. (d) \dpolar($16,8,\ell=4)$ : Scaling the \dpolar encoding to higher-order kernels enables us to achieve good reliability. We are the first to explore this design space}
\label{fig:architecture}
\end{center}
\vskip -0.2in
\end{figure*}

Polar codes, introduced by Erdal Arıkan \cite{arikan2009channel}, are the first deterministic code construction to achieve the Shannon capacity while maintaining low encoding and decoding complexity. This section formally defines Polar codes and motivates our method.


\textbf{Polar Encoding. }
A Polar code can be described by Polar($n$,$k$,$\mathcal{F}$). Here, $n$ is the block length; $n = 2^m$ for some integer $m$, $k$ is the number of information bits, and $\mathcal{F}$ represents the set of ``frozen" bit positions. Typically, the positions corresponding to the noisiest $n-k$ bit channels arising due to polarization are chosen to be frozen. A polar encoder maps information bits $\bu \in \{0,1\}^k$ to a binary codeword $\bx \in \{0,1\}^n$. The basic building block of Polar codes is the Plotkin transform: $\{0,1\}^d \times \{0,1\}^d \to \{0,1\}^{2d}$. This mapping for a pair of input bits $(u, v)$ can be represented by the matrix $G_2 = \begin{bmatrix} 1 & 0 \\ 1 & 1 \end{bmatrix}$, transforming $(u_0, u_1)$ into $(u_0 \oplus u_1, u_1)$, where $\oplus$ denotes the XOR operation. Consistent with the coding theory literature, we term such a building block a \textit{kernel}. The encoding matrix for block length $n = 2^m$ is obtained by taking the \textit{Kronecker product} of the base kernel $G_2$ $m$ times.


Utilizing this structure, the encoding can be efficiently performed via a recursive coordinate-wise application of the Plotkin transform on a binary tree, called the \textit{Plotkin tree}. To encode a block of message bits $\bu= (u_0,\ldots, u_{k-1}) \in \binary^k$, we first embed them into a source message vector $\bma \define (m_1,\ldots,m_n) = (0, \ldots, u_0, 0,  \ldots, u_1, 0, \ldots, u_{k-1}, 0, \ldots) \in \binary^n$, where $\bma_{\mathcal{I}}= \bu$ and $\bma_{\mathcal{I}^C}=0$ for some $\mathcal{I} \subseteq [n]$.  Since the message block $\bma$ contains the information bits $\bu$ only at the indices pertaining to $\mathcal{I}$, the set $\mathcal{I}$ is called the \emph{information set}, and its complement $\mathcal{F} = \mathcal{I}^C$ the \emph{frozen set}. We describe the encoding on a Plotkin tree via a small example, Polar(4,3) - illustrated in ~\prettyref{fig:architecture}(a). Here, $\mathcal{F}$=\{0\}.  Consider an input of size $k=3$, $u = [u_0, u_1, u_2]$. At the input level (depth 1), we freeze $m_0$, i.e., $m_0 = 0$  and assign $\bu$ to the remaining positions. Applying the Plotkin transform, $(0, u_0) \to (0 \oplus u_0, u_0)$ and $(u_1, u_2) \to (u_1 \oplus u_2, u_2)$. At the second level, we apply the same operation \textit{coordinatewise} to these vectors \textit{i.e,} $((0 \oplus u_0), (u_1 \oplus u_2)) \to ((0 \oplus u_0) \oplus (u_1 \oplus u_2), u_1 \oplus u_2)$ and $(u_0, u_2) \to (u_0 \oplus u_2, u_2)$. The final encoded vector is the concatenation of the outputs from the second-level nodes \textit{i.e,} $(u_0 \oplus u_1 \oplus u_2, u_0 \oplus u_2, u_1 \oplus u_2, u_2)$. For a general $(n,k)$ polar code, the encoding proceeds similarly up to $m = \log_2{n}$ levels.

\textbf{Polar decoding.} The encoded messages are corrupted by a noisy channel $W \define W_{Y|X}$. The successive cancellation (SC) algorithm is one of the most efficient decoders for Polar codes and is optimal asymptotically. The basic principle behind the SC algorithm is to sequentially decode each message bit $u_i$ according to the conditional likelihood given the corrupted codeword $y$ and previously decoded bits $\hat{\bu}^{(i-1)}$. LLR for $i^{\text{th}}$ bit can be computed as
\begin{equation}
    L_i = \log \left( \frac{\mathbb{P}(u_i = 0| y, \hat{\bu}^{(i-1)} )}{\mathbb{P}(u_i = 1| (y, \hat{\bu}^{(i-1)} )} \right).
\end{equation}

SC decoding is described in detail in ~\prettyref{app:sc_decoder}.  

\textbf{Large kernel Polar codes.}  
Polar codes are optimal (capacity-achieving) at asymptotic blocklengths due to the phenomenon of channel polarization. \cite{korada2010polar} prove that replacing the conventional $2 \times 2$ kernel with an $\ell \times \ell$ binary kernel ($\ell>2$) still results in polarization, provided this matrix is non-singular and not upper triangular under any column permutation. Further, large kernel polar codes that achieve capacity at shorter block lengths compared to conventional polar codes have been found, as indicated by better \textit{scaling exponents} \cite{fazeli2014scaling, fazeli2020binary}. Notably, the kernel size must be expanded to 8 to surpass Arikan's $G^{\otimes 2}$ kernel, as no size 4 linear kernel offers an improved scaling exponent. These kernels enable better finite-length performance at the cost of increased decoding complexity.

Large kernel polar encoding and decoding proceeds similarly to conventional Polar codes via an $\ell \times \ell$ kernel. An exemplar kernel with $\ell=4$ is simply the Kronecker product of $G$ with itself, $G_4 = G \otimes G$. This is a Plotkin transform with 4 inputs, referred to as \textit{Plotkin-4}. We illustrate this via an example of a $(n=16,k=8)$ code with kernel size $\ell=4$ in \prettyref{fig:architecture} . In this example, the message $\bu$ is input at the information positions $\mathcal{I} = \mathcal{F}^C = \{7,9,10,11,12,13,14,15\}$. The remaining positions are frozen to 0. The kernel is applied in parallel to groups of $\ell=4$ bits. Mirroring the conventional polar encoding, we iteratively follow this process at each level of the tree. At the second level, we apply Plotkin-4 coordinatewise to the vectors $(v_{10}, v_{11}, v_{12}, v_{13})$ to obtain the codeword $x$.

In this work, we design a non-linear generalization of the polar coding and decoding structure by both expanding the kernel size and parameterizing them as neural networks. 
By expanding the design space to include non-linearity and larger kernels, we aim to discover more reliable codes within the neural Plotkin code family.


\section{\dpolar codes}
\subsection{Proposed Architecture}
We design \dpolar codes by generalizing the encoding and decoding structures of large kernel Polar codes. A \dpolar$(n,k,\ell)$ code maps bits $\bu \in \{0,1\}^k$ to a codeword $\bx \in \mathbb{R}^n$, using the neural Plotkin tree of kernel size $\ell$.

\begin{figure*}[ht]
\captionsetup[subfloat]{labelformat=simple, labelsep=none, listofformat=subsimple}
\renewcommand{\thesubfigure}{\alph{subfigure})}
\centerline{
\subfloat[Stage 1 - Curriculum learning for kernels]
{
 \includegraphics[width=0.5\linewidth, trim={0cm 5cm 0cm 1cm}, clip]{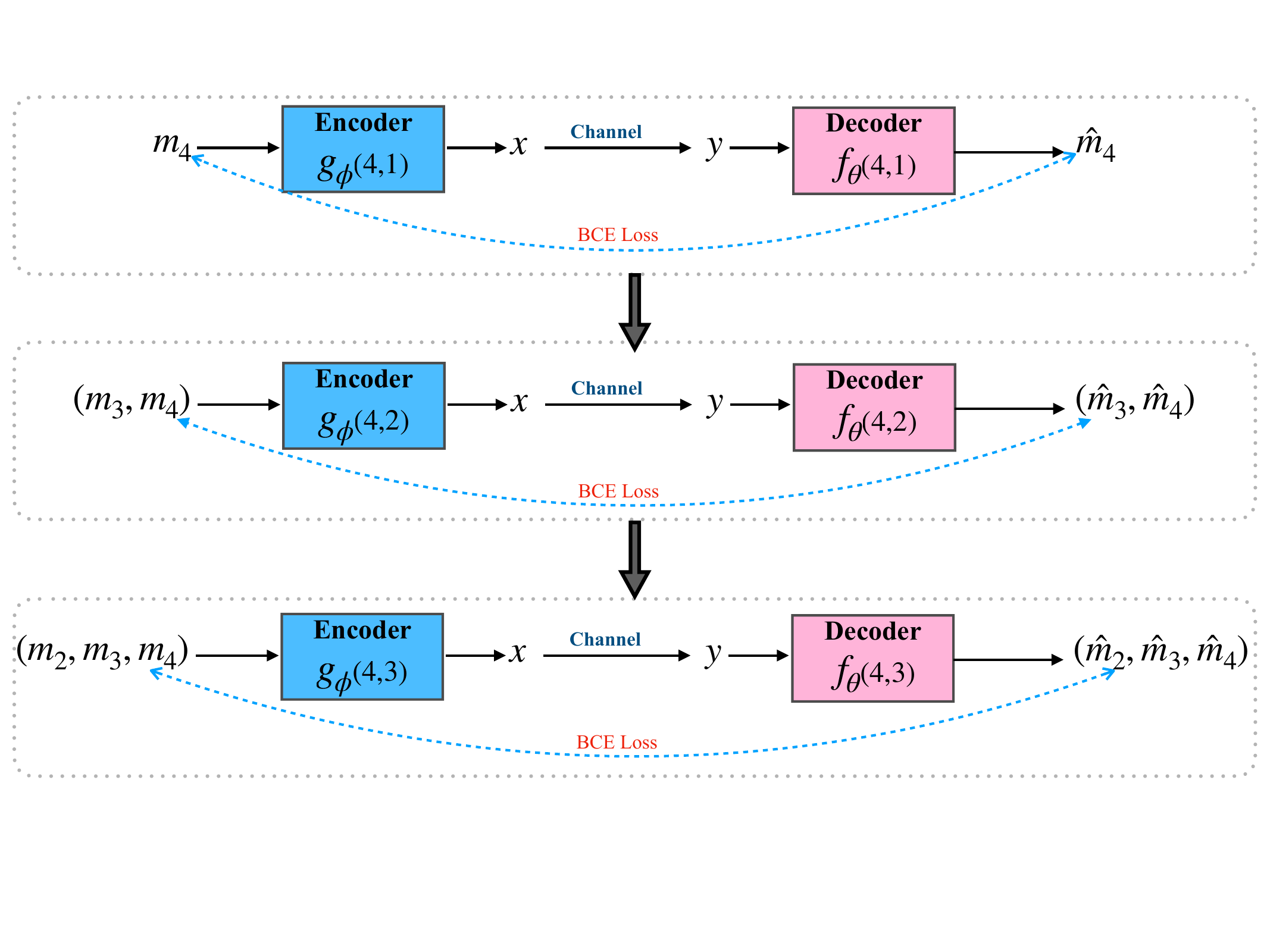}
  \label{fig:curriculum_a}
}
\hfill
\subfloat[Stage 2 - Initialize encoder and decoder by pretrained kernels]
{
 \includegraphics[width=0.5\linewidth, trim={5cm 8cm 6cm 3cm}, clip]{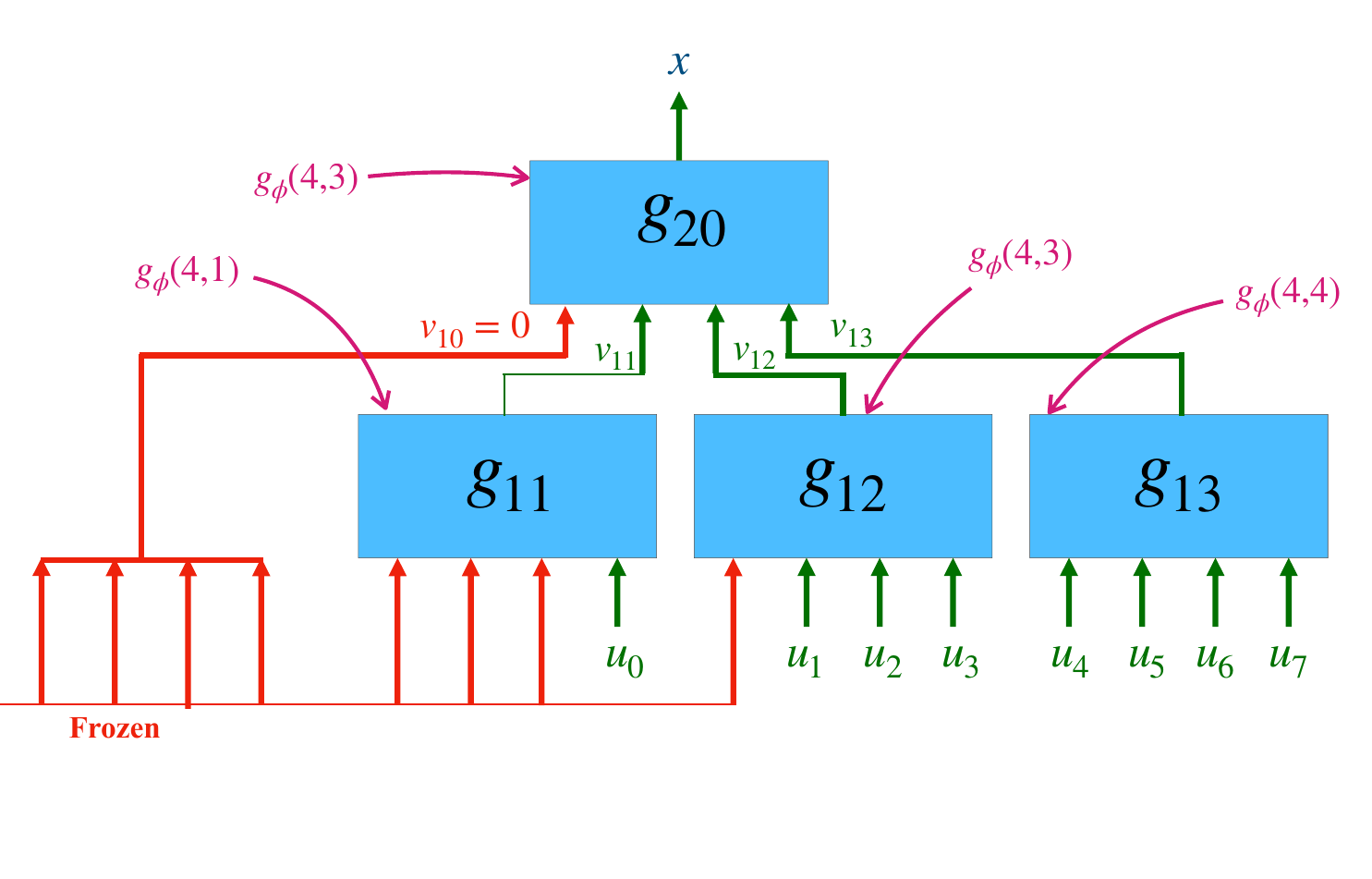}
  \label{fig:curriculum_b}
}
}
\caption{The curriculum to train \dpolar$(16,8,\ell=4)$ proceeds in two phases: (a) In Phase 1, the kernels $(4,i)$ are trained progressively for $i = 1,\cdots, \ell$. (b) In phase 2, we initialize each kernel in the encoder by the respective kernels (and similarly for the decoders).  
For instance, the kernel $g_{11}$ has one information input bit; we initialize it with a pretrained $g_\phi(4,1)$. Similarly, $g_{20}$ has three information input bit-groups; we initialize it with a pretrained $g_\phi(4,3)$.}
\label{fig:curriculum}
\vspace{-1em}
\end{figure*}

\textbf{\dpolar encoder.} Building on the foundational encoding structure of Polar codes (\prettyref{fig:architecture}a), which employs the recursive application of a binary kernel over a Plotkin tree, we introduce \dpolar codes as a non-linear extension of this concept. The first major modification in our approach is expanding the conventional 
$2 \times 2$ kernel to a larger $\ell \times \ell$ kernel. 
This is inspired by the existence of larger binary kernels that outperform the Arikan kernel (\prettyref{app:large_polar}.)

On the other hand, we generalize the Plotkin transform by replacing each XOR operation with a 2-input NN $g_{d,b}$, where $d$ is the depth in the encoding tree, and $b$ is the index of the NN at depth $d$.
A \dpolar$(n,k,\ell)$ inherits the same parameters as a Polar$(n,k)$ code.
The encoding structure is analogous to conventional polar codes and is illustrated in \prettyref{fig:architecture}(c) - \dpolar(4,3) with $\ell=2$.
We use the same frozen set $\mathcal{F}=\{0\}$ as the conventional polar code. Applying the neural Plotkin transform, $(0, u_0) \to v_{10} = (g_{10}(0, u_0), u_0)$ and $(u_1, u_2) \to v_{11} = (g_{11}(u_1,u_2), u_2)$. At the second level, we apply the kernel $g_{20}$ \textit{coordinatewise} to these vectors.  This coordinatewise operation is the key inductive bias in \dpolar encoding. The final encoded vector is the concatenation of the outputs from the second-level nodes \textit{i.e,} $(g_{20}(v_{10}, v_{11}), v_{11})$. For a general \dpolar$(n,k,\ell=2)$code, the encoding proceeds in a similar way upto $m = \log_{2}{n}$ levels. 


 Our work synergizes these two directions by employing a non-linear kernel of size $\ell \times \ell$. \prettyref{fig:architecture}(d) illustrates \dpolar$(16,8,\ell=4)$, which mirrors the Plotkin-4 encoder for Polar$(16,8)$. The encoding proceeds similarly, with neural networks $g_{d,b}$ replacing the Plotkin mapping. A power constraint, $\|x\|^2 = n$ is enforced at the output of the encoder. \prettyref{app:arch} discusses the \dpolar encoder architecture in detail. 


\dpolar leverages the Polar encoding structure by parameterizing the kernel at each internal node at depth $d$ and bit position $b$ of the Plotkin tree by a neural network $g_{d,b}$. We hypothesize that this innovation unlocks performance gains through the expanded function space afforded by both non-linearization and the increased kernel size. Additionally, this formulation allows us to scale neural Polar codes to larger blocklengths and rates.

\textbf{\dpolar-SC decoder.}
To decode the received codewords corrupted by noise, we design the \dpolar-SC decoder, a neural generalization of SC decoding. The \dpolar-SC decoder consists of a decoding tree, consisting of $\ell$ decoding networks $f_{d,\ell b+j}, j:0 \to \ell -1$, matching every kernel $g_{d,b}$. Notably, each decoding NN $f_{d,\ell b+j}$ is applied coordinatewise and takes as input the LLR from the previous $j-1$ outputs, along with the $\ell$ inputs to the kernel. \dpolar-SC is detailed in ~\prettyref{app:ell_decoder}. 


\begin{figure*}[ht]
\captionsetup[subfloat]{labelformat=simple, labelsep=none, listofformat=subsimple}
\renewcommand{\thesubfigure}{\alph{subfigure})}
\centerline{
\subfloat[ $n=256,k=37$]
{  \hspace*{-0.25in}
 \includegraphics[width=0.33\linewidth]{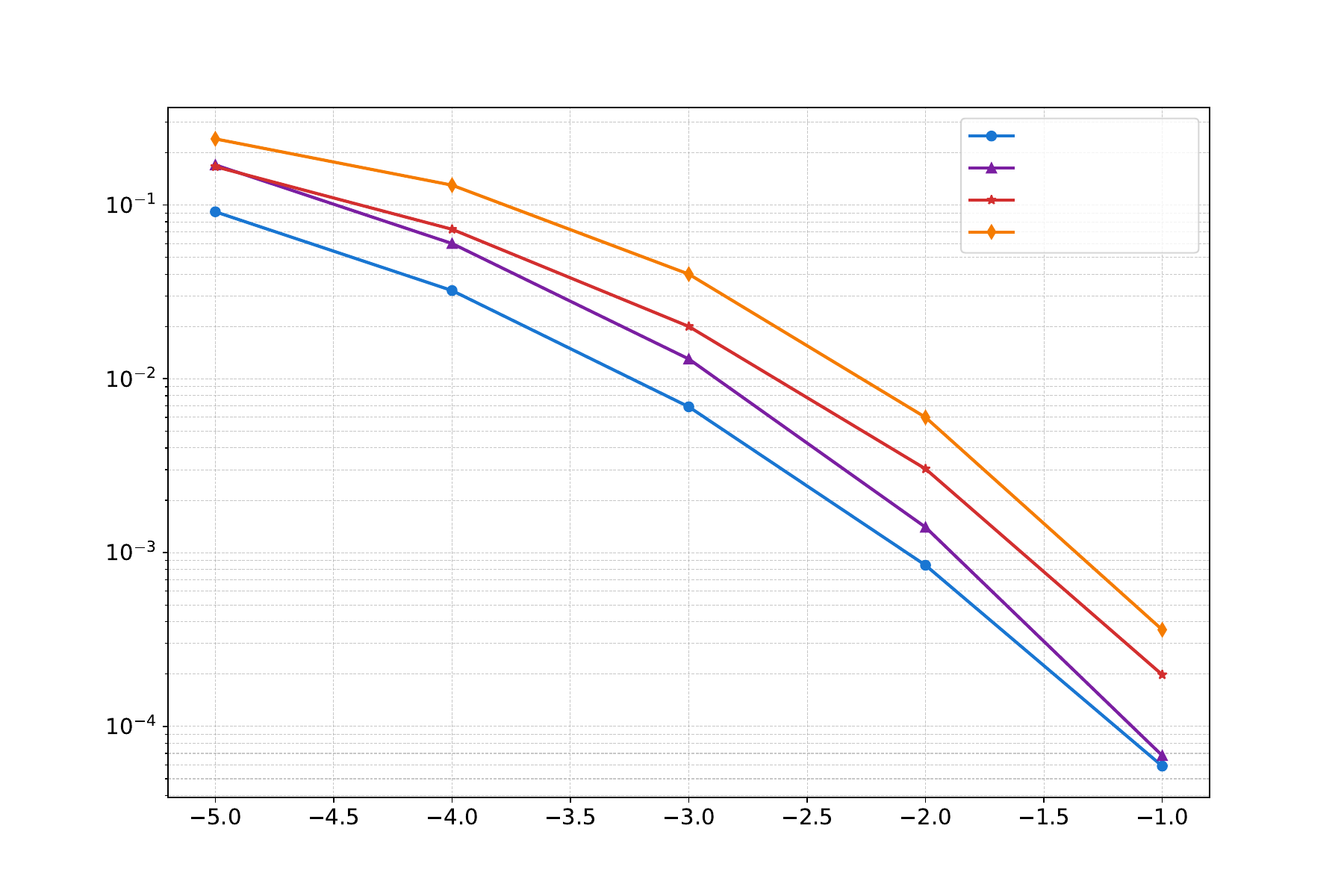} 

    \put(-35,90){\fontsize{3}{5}\selectfont \dpolar}
    \put(-35,86.5){\fontsize{3}{5}\selectfont KO(8,2)}
    \put(-35, 83){\fontsize{3}{5}\selectfont Polar(256,37)}
    \put(-35,78){\fontsize{3}{5}\selectfont RM(8,2)}
    \put(-120,0){\fontsize{7}{5}\selectfont Signal-to-noise ratio (SNR) [dB]}
    \put(-155,50){\rotatebox[origin=t]{90}{\fontsize{7}{5}\selectfont Bit Error Rate}}
  \label{fig:25637}
 }
\hfill
\subfloat[ $n=256,k=64$]
{
 \includegraphics[width=0.33\linewidth]{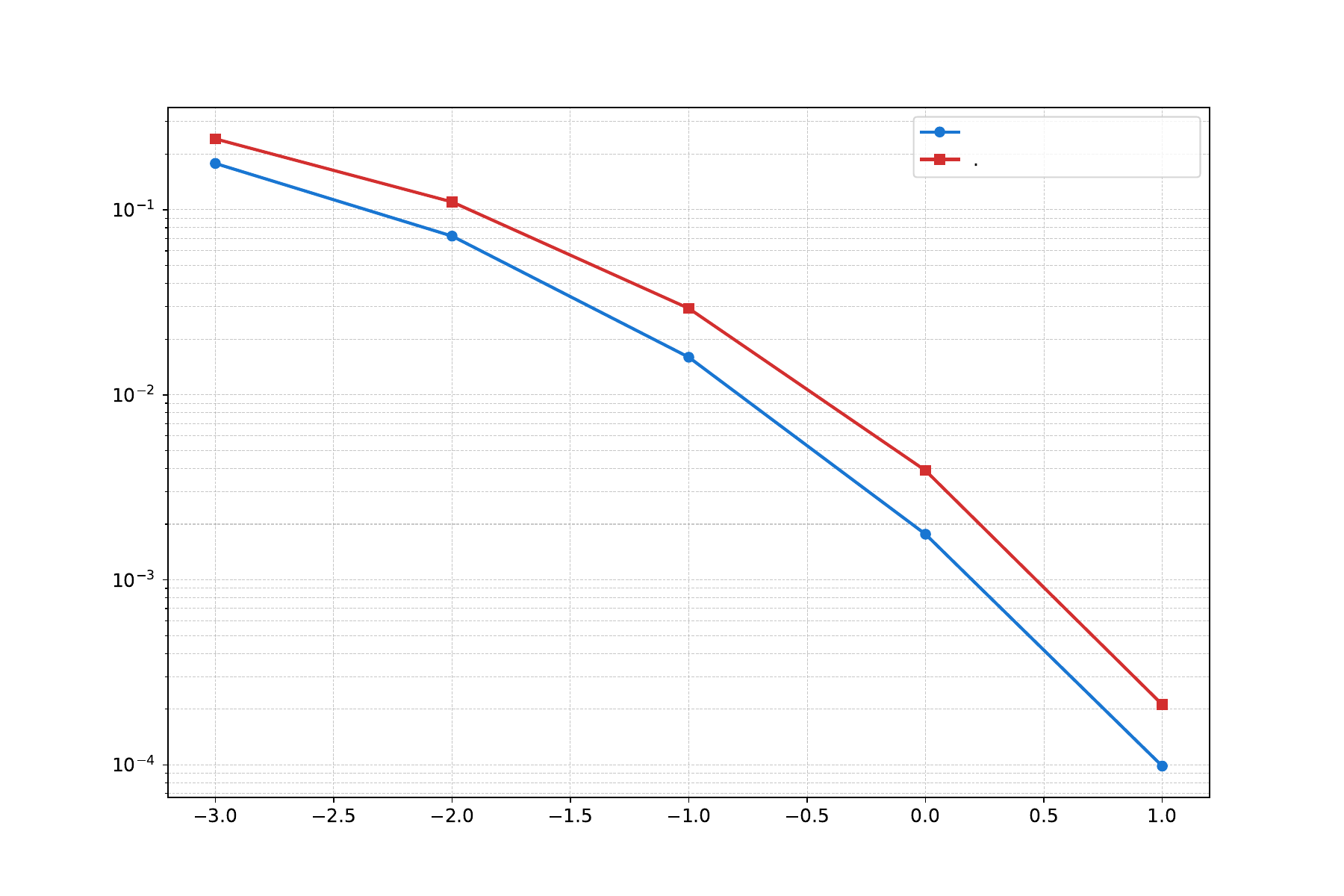}
    \put(-40,91){\fontsize{3}{5}\selectfont \dpolar}
    \put(-40,87.5){\fontsize{3}{5}\selectfont Polar}
    \put(-120,0){\fontsize{7}{5}\selectfont Signal-to-noise ratio (SNR) [dB]}
    \put(-155,50){\rotatebox[origin=t]{90}{\fontsize{7}{5}\selectfont Bit Error Rate}}
  \label{fig:25664}
}
\hfill
\subfloat[ $n=256,k=28$]
{
 \includegraphics[width=0.33\linewidth]{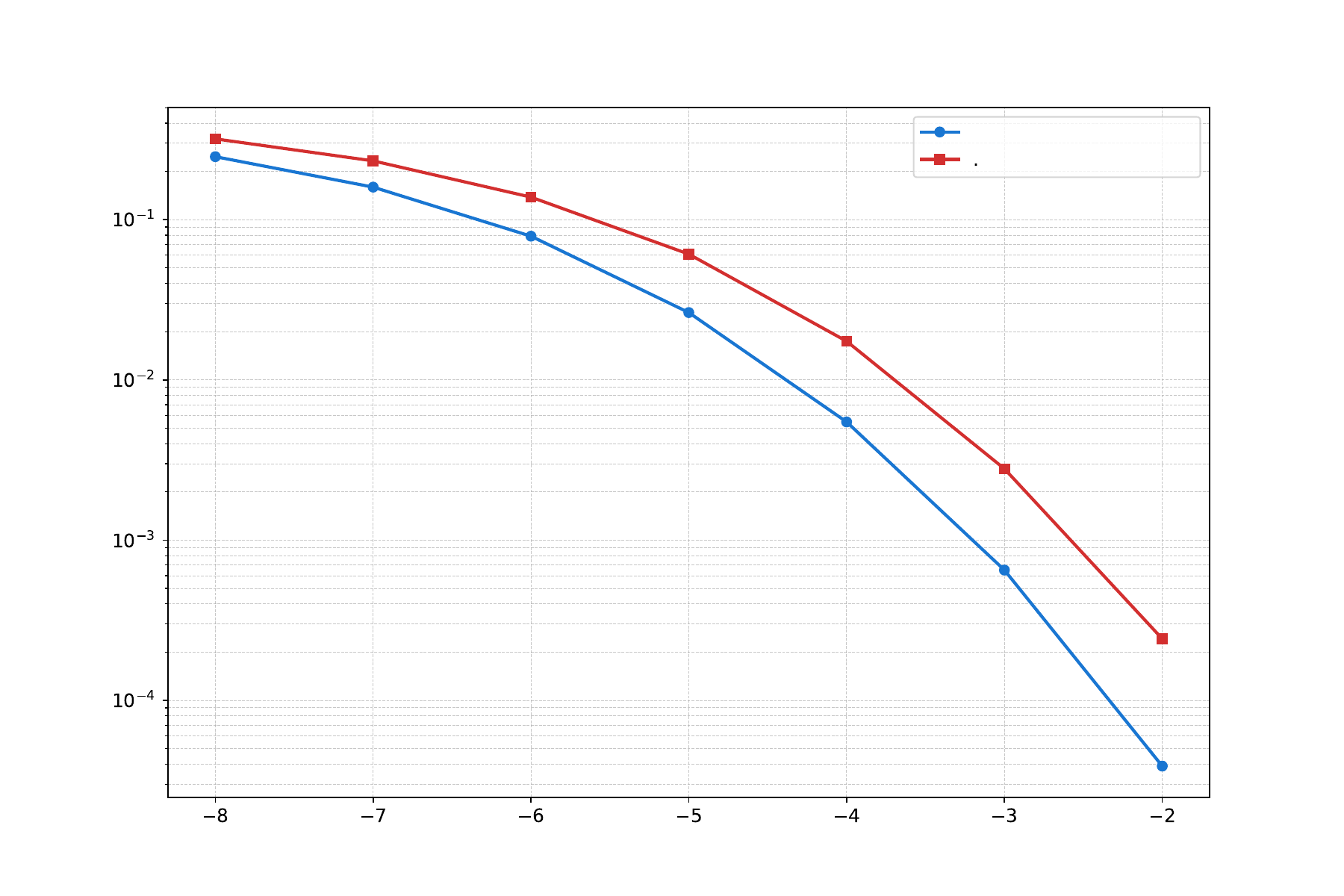}
    \put(-40,91){\fontsize{3}{5}\selectfont \dpolar}
    \put(-40,87.5){\fontsize{3}{5}\selectfont Polar}
    \put(-120,0){\fontsize{7}{5}\selectfont Signal-to-noise ratio (SNR) [dB]}
    \put(-155,50){\rotatebox[origin=t]{90}{\fontsize{7}{5}\selectfont Bit Error Rate}}
  \label{fig:25628}
}
 
}
\caption{(a) \dpolar improves over state-of-the-art KO codes \cite{Makkuva2021}, and RM, Polar codes at $n$=256,$k$=37. (b,c) \dpolar 
is suitable for a variety of rates and retains gains over Polar, while KO is not suitable for these rates.}
\label{fig:awgn_restplots}
\vspace{-1em}
\end{figure*}
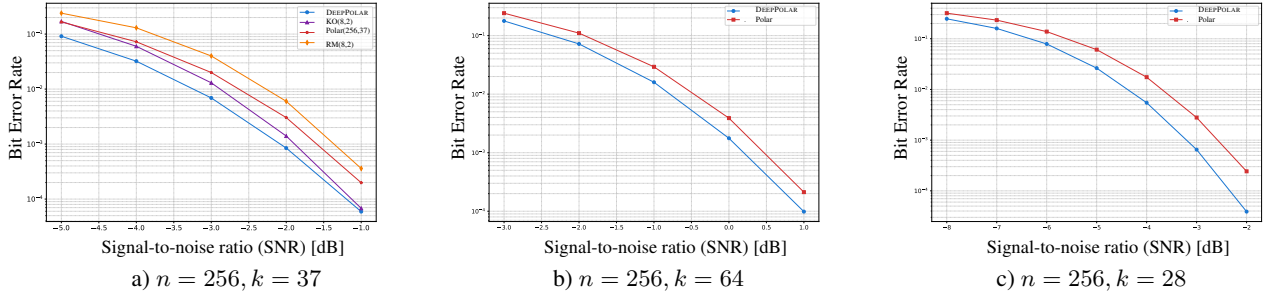

\subsection{Training methodology : Curriculum learning.}\label{sec:curriculum}
The primary challenge in training neural channel codes is the astronomical space of potential codewords; the task is to find a noise-robust mapping for each binary string in the $k$-dimensional Boolean hypercube to a codeword in $\mathbb{R}^n$. This challenge is pronounced even in the short-to-medium blocklength regime we focus on, for instance, training a (256,37) code involves designing $2^{37}$ codewords in a 256-dimensional space. During training, we encounter a tiny fraction of the message space $(<1\%)$, making the NN's ability to effectively generalize to unseen messages a pivotal factor in learning effective codes. While principled architectural choices are vital for generalization, the training methodology is equally important. Typically, an end-to-end training strategy would involve minimizing the binary cross entropy loss between the actual message bits and estimated message bits :
\begin{equation}
\mathcal{L} = \!=\!-\sum_{i=1}^{k} \left[ u_i \log(\hat{u}_i) + (1\!-\!u_i) \log(1\!-\!\hat{u}_i) \right] 
\end{equation}
However, as highlighted in \prettyref{sec:ablation}, direct training often does not generalize well to high SNR scenarios, characterized by infrequent error events.

\textbf{Curriculum.} We address this by introducing a principled two-stage curriculum training procedure that capitalizes on the inherent nested hierarchy of Polar codes. We leverage two key properties of the Polar coding structure: \textit{(1)Hierarchy in k}: A Polar$(n,k)$ subsumes all codewords of lower-rate subcodes $\polar(n,i), 1 \leq i \leq k$. Curriculum training leveraging the hierarchy in $k$ improved both convergence speed and reliability in the context of neural polar decoding \cite{hebbar2023crisp}. In the first stage, we use a similar C2N curriculum to pretrain the parameters for each kernel. We progressively train encoder-decoder pairs for \dpolar($n=\ell, k=j, \ell)$ codes, for $j=1 \to k$ as highlighted in \prettyref{fig:curriculum_a}), denoted by $g_\phi(\ell, j)$, $f_\theta(\ell, j)$. 
\textit{(2)Hierarchy in n}: In the Plotkin tree, an $\ell \times \ell$ kernel is applied coordinatewise at each level of the Plotkin tree. In the second stage of our curriculum, we initialize each kernel in \dpolar($n$,$k$) encoder and decoder with the pre-trained kernels from the first stage, applied at all depths, as illustrated in \prettyref{fig:curriculum_b}. 
For instance, $g_{12}$ and $g_{20}$, which have three non-frozen inputs, are initialized by a pre-trained kernel $g_\phi(4,3)$. 
Such a systematic and phased training approach significantly enhances the network's learning efficiency and generalizability (\prettyref{sec:ablation}).

\section{Main Results}

\subsection{Data generation.}
We generate synthetic input data for the encoder by randomly sampling from a boolean hypercube \textit{i.e,} $\{0,1\}^k$. A randomly sampled white Gaussian noise is added to the output of the encoder. During training, the variance of the Gaussian noise is carefully chosen based on the length and rate of the desired channel code. 

\subsection{Baselines.} 
\textcolor{black}{\dpolar extends conventional Polar encoding and decoding structures using neural networks. We evaluate its performance against state-of-the-art learning-based codes, KO codes \cite{Makkuva2021}. Similar to \dpolar, KO codes replaces elements of Reed-Muller by neural networks to enhance error correction capability. We also compare \dpolar to traditional Polar codes employing Arikan's kernel and decoded via the successive cancellation (SC) algorithm. Both codes utilize the same frozen positions, determined using the Tal-Vardy method \cite{tal2013construct}. We further discuss the impact of frozen position selection in \prettyref{app:5g}. \prettyref{app:large_polar} discusses and compares \dpolar to recent large-kernel polar code constructions.}


\subsection{Results}
\textbf{\dpolar codes outperform Polar codes.} As highlighted in \prettyref{fig:awgn_restplots}, \dpolar codes achieve enhanced performance over traditional Polar codes with SC decoding and Reed-Muller codes with Dumer decoding across a broad range of SNRs in the presence of additive white Gaussian noise (AWGN), in terms of bit error rate (BER). Notably, \dpolar(256,37,$\ell=16$) outperforms KO($8,2$) (\prettyref{fig:25637}), the state-of-the-art neural code {mapping $37$ information bits to a length-$256$ codeword} \cite{Makkuva2021} in the short-to-medium blocklength regime. \dpolar codes, owing to their Polar-like encoding framework, accommodate a much wider range of rates than KO codes that rely on the algebraic structure of RM codes. This versatility is illustrated in \prettyref{fig:25664} and \prettyref{fig:25628}, where \dpolar codes outperform Polar codes across diverse rates and SNRs (cf. KO codes and RM codes do not exist for these rates).  
Further, \dpolar{} is robust to non-AWGN deviations, achieving gains over Polar codes on fading channels and radar noise (\prettyref{app:robustness}).


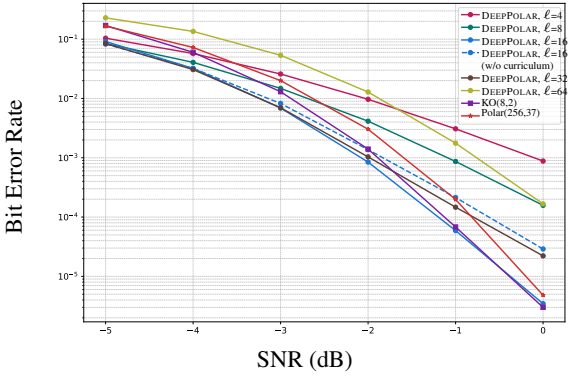
\begin{figure}[ht]
\vskip -0.1in
\begin{center}
\includegraphics[width=\columnwidth]{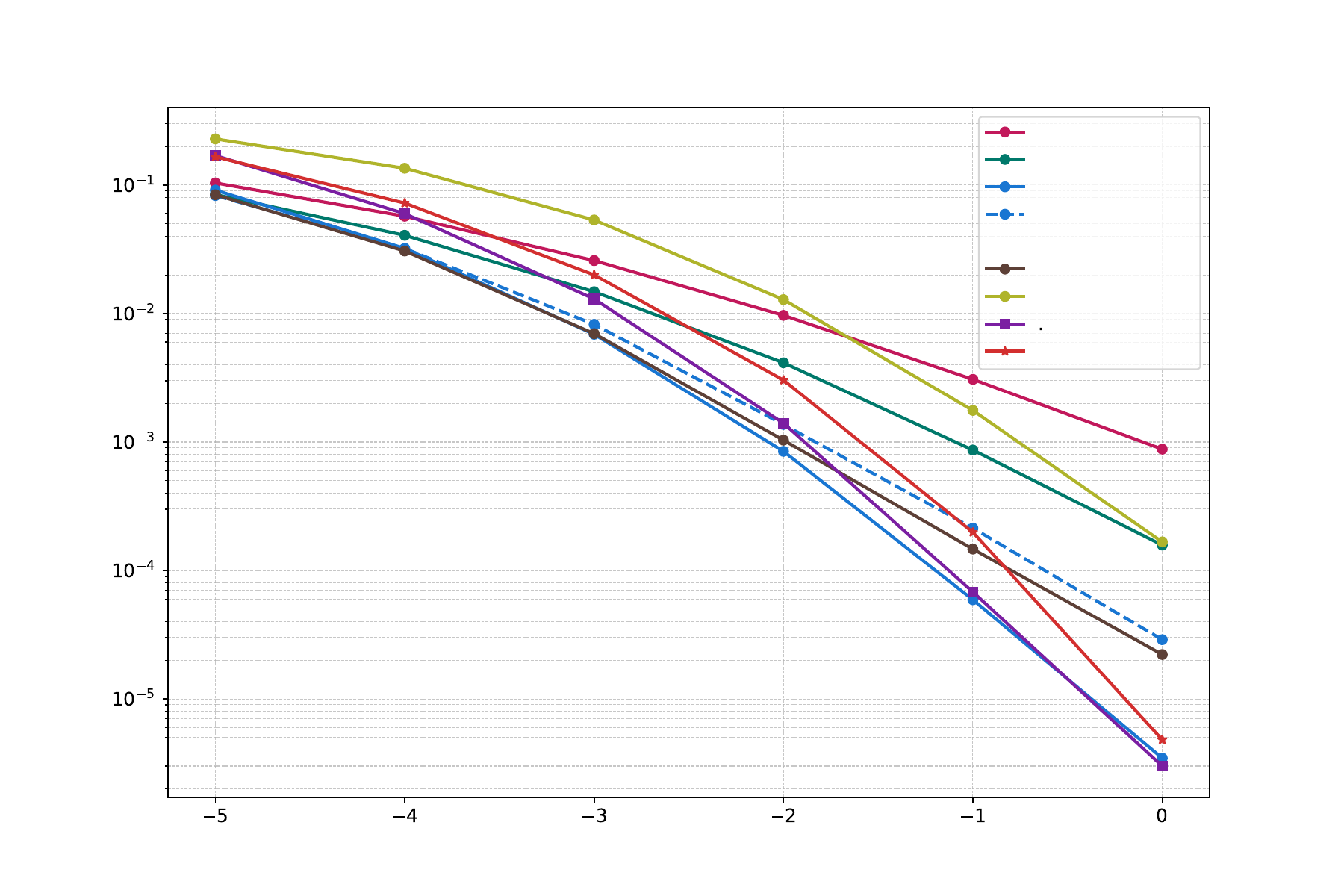}
\put(-55,132){\fontsize{4}{5}\selectfont \dpolar, $\ell$=4}
\put(-55,127){\fontsize{4}{5}\selectfont \dpolar, $\ell$=8}
\put(-55,122){\fontsize{4}{5}\selectfont \dpolar, $\ell$=16}
\put(-55,117.5){\fontsize{4}{5}\selectfont \dpolar, $\ell$=16}
\put(-55,113){\fontsize{4}{5}\selectfont (w/o curriculum)}
\put(-55,108){\fontsize{4}{5}\selectfont \dpolar, $\ell$=32}
\put(-55,103){\fontsize{4}{5}\selectfont \dpolar, $\ell$=64}
\put(-55,99){\fontsize{4}{5}\selectfont KO(8,2)}
\put(-55,95){\fontsize{4}{5}\selectfont Polar(256,37)}
\put(-140,0){\footnotesize SNR (dB)}
\put(-235,70){\rotatebox[origin=t]{90}{\footnotesize Bit Error Rate}}
\caption{$n$=256,$k$=37 : Reliability of \dpolar improves as the kernel size $\ell$ is increased. We find that \dpolar with $\ell$=16,32 beats KO codes in BER. Moreover, curriculum training is crucial to improve reliability at high SNRs.} 
\label{fig:kernel_size_curriculum}
\end{center}
\end{figure}


\textbf{Effect of kernel size.} 
\prettyref{fig:kernel_size_curriculum} highlights that scaling the kernel size $\ell$ is crucial for effective training and \dpolar outperforming baselines. In case $n$ is not an integer exponent of $\ell$ - i.e., $\ell^{m}\leq n < \ell^{m+1}$ and $\ell^{m}$ divides $n$ then the kernel size at the root is $n/\ell^m$. An increase in kernel size expands the function space the encoder can represent, facilitating learning more robust representations. However, empirical experiments reveal that using $\ell = \sqrt{n}$ , which results in an encoding and decoding tree of depth 2, is a good heuristic. Indeed, \prettyref{fig:kernel_size} highlights significant improvements in performance upto $\ell=16,32$, whereas further scaling $\ell=64$ leads to a decline. This trend indicates a bias-variance tradeoff, suggesting the ideal kernel size strikes a balance between model complexity and its generalization capabilities.

\subsection{Interpretation}

\begin{figure}[ht]
\begin{center}
\includegraphics[width=0.7\columnwidth]{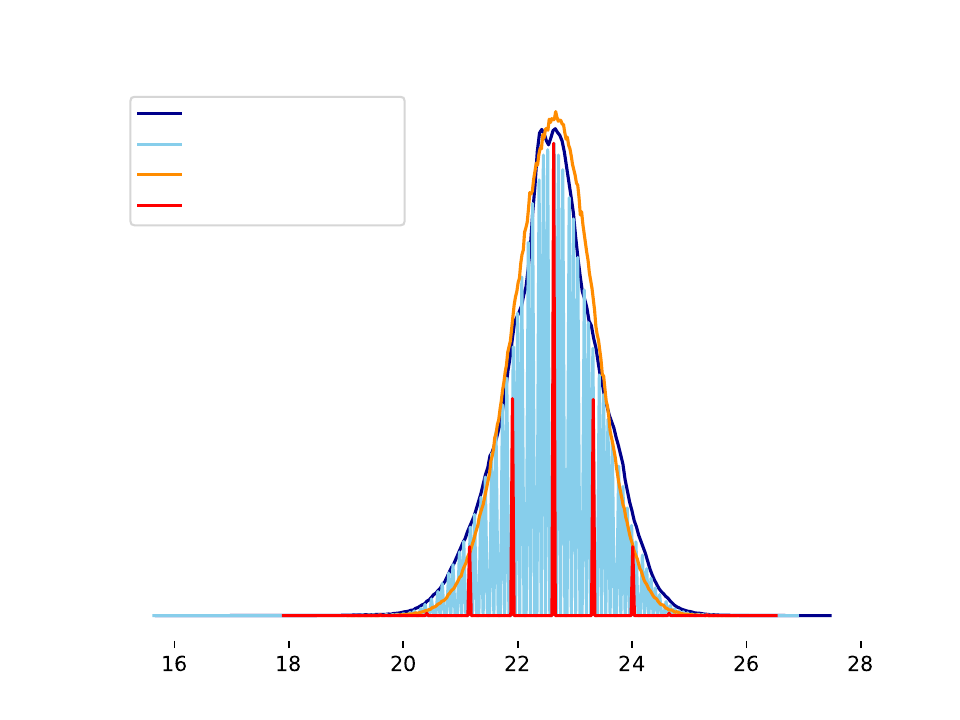}
\put(-132,102){\fontsize{4}{5}\selectfont \dpolar}
\put(-132,97){\fontsize{4}{5}\selectfont \dpolar-binary}
\put(-132,92){\fontsize{4}{5}\selectfont Gaussian}
\put(-132,87){\fontsize{4}{5}\selectfont Polar}
\put(-100,0){\footnotesize Pairwise distance}

\caption{(n=256,k=37,$\ell=16$): Unlike Polar codes, the distribution of pairwise distribution between codewords of \dpolar shows a strong resemblance to the Gaussian codebook. \dpolar-binary code retains a Gaussian-like distance profile. (pdfs and pmfs have been renormalized for better visualization.)} 
\label{fig:distance}
\end{center}
\vspace{-1em}
\end{figure}
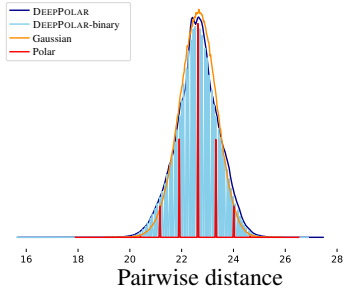

\textbf{Interpreting the encoder.} To interpret the encoder, we examine the distribution of pairwise distances between codewords (\prettyref{fig:distance}). Gaussian codebooks achieve capacity and are optimal asymptotically ~\cite{shannon1948mathematical}. Remarkably, the distribution of \dpolar codewords closely resembles that of the Gaussian codebook. This surprising phenomenon has also been observed in the closely related KO codes and is a testament to the potential of the marriage between efficient algebraic code structures and DL.



        

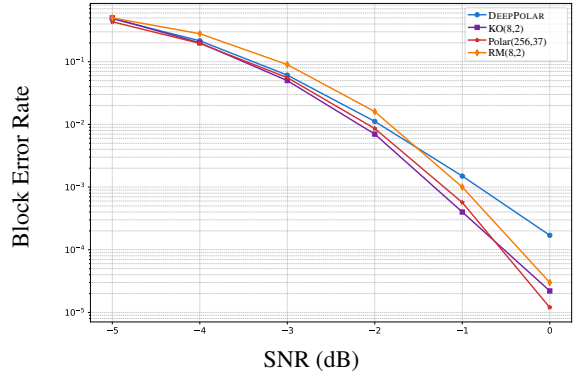
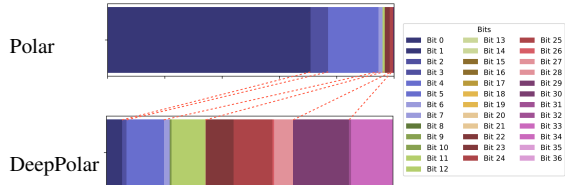
\begin{figure}[ht]
\captionsetup[subfloat]{labelformat=simple, labelsep=none, listofformat=subsimple}
\renewcommand{\thesubfigure}{\alph{subfigure})}
\subfloat[ BLER performance ($n$=256,$k$=37)]{%
  \includegraphics[clip,width=\columnwidth]{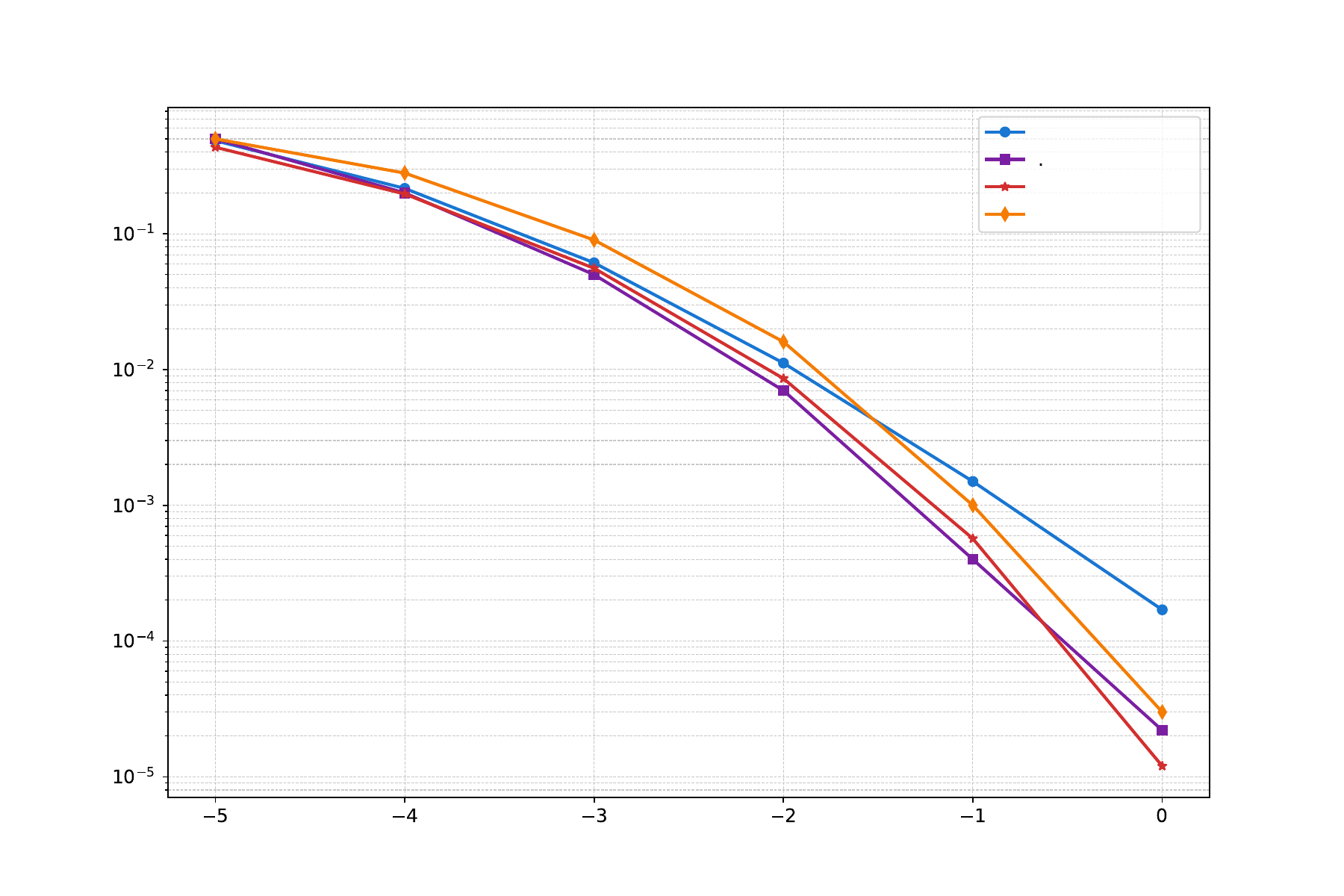}%
\put(-55,132){\fontsize{4}{5}\selectfont \dpolar}
\put(-55,127){\fontsize{4}{5}\selectfont KO(8,2)}
\put(-55,122){\fontsize{4}{5}\selectfont Polar(256,37)}
\put(-55,118){\fontsize{4}{5}\selectfont RM(8,2)}
\put(-140,0){\footnotesize SNR (dB)}
\put(-235,70){\rotatebox[origin=t]{90}{\footnotesize Block Error Rate}}
\label{fig:bler_256_37}
}
\vspace{-1mm}
\subfloat[ Distribution of Errors ($n$=256,$k$=37)]{%
  \includegraphics[trim={0 10cm 0 4cm},clip,width=0.9\columnwidth]
  {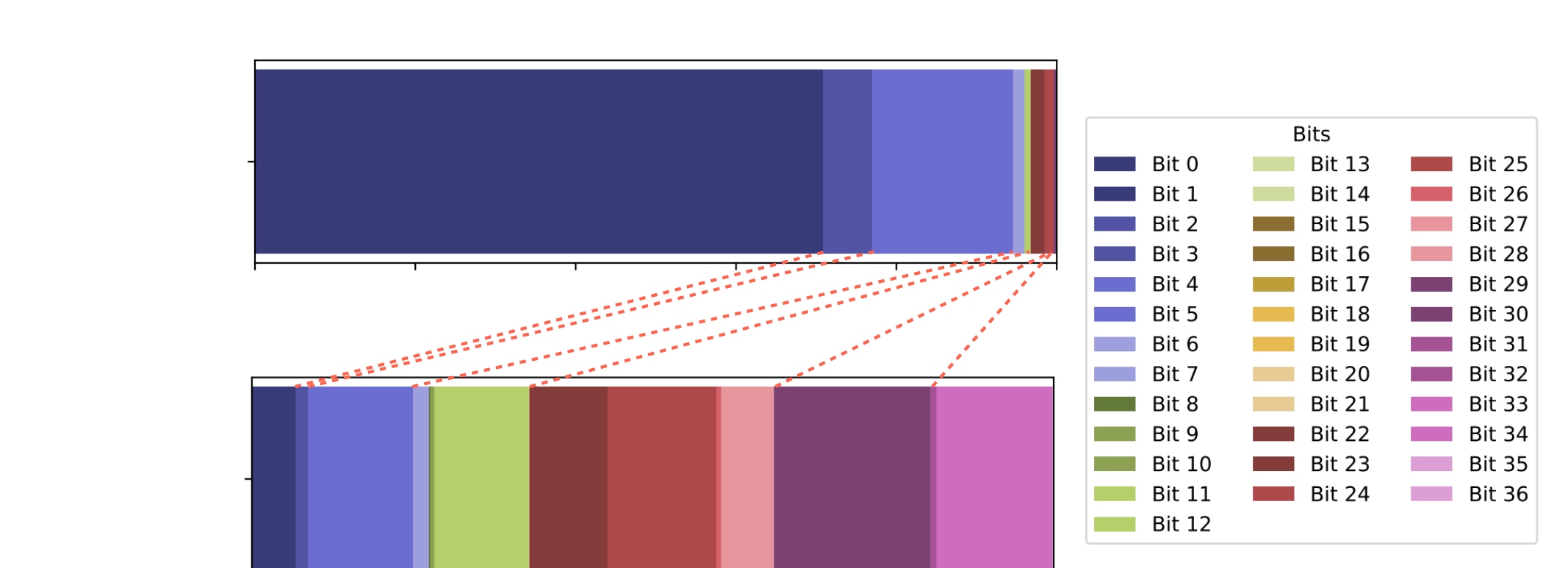}%
  \put(-214,50){\fontsize{8}{5}\selectfont Polar}
  \put(-214,5){\fontsize{8}{5}\selectfont DeepPolar}
  \label{fig:error_analysis}
}
\caption{a) The block error rate performance of \dpolar codes is subpar. b) The distribution of the first error occurrence across bit positions reveals that this tradeoff is a consequence of a more uniform error distribution, driven by the use of BCE loss, which is a surrogate for BER.}
\vspace{-1em}
\end{figure}

\textbf{Interpreting the decoder.} In our study, the training loss used is binary cross entropy (BCE), which serves as a surrogate for the bit error rate (BER). However, it is important to recognize that optimizing for BER does not necessarily translate to improved block error rate (BLER), an important figure of merit in practical systems. Our observations, as indicated in \prettyref{fig:bler_256_37}, show that \dpolar underperforms in BLER compared to baseline methods {(See \prettyref{app:additional} for more results)}. The BLER, defined as $\mathbb{P}[\hat{\bu} \ne \bu]$, can be expressed as the cumulative probability of bit errors occurring at each position when no errors were made in the previous ones:
\begin{equation}
    \mathbb{P}[\hat{\bu} \ne \bu] = \sum_{i = 1}^k \mathbb{P}[\hat{u}_i \ne u_i | \hat{\bu}_{1:i-1} = \bu_{1:i-1}]
\end{equation}
An analysis of error distribution (\prettyref{fig:error_analysis}) reveals notable differences between SC and \dpolar-SC decoders despite both employing a sequential decoding approach.  Due to the effect of channel polarization, most errors in the polar decoding tend to occur at bit position 0, which predominantly drives the BLER. In contrast, errors in \dpolar decoding are more evenly distributed across various bit positions. This uniformity is a consequence of the BCE loss prioritizing BER over BLER. This trade-off leads to a marginal decrease in BLER performance. Identifying a stable loss function directly targeting BLER, is an interesting direction for future research.

\subsection{Ablation studies}\label{sec:ablation}

\textbf{\dpolar-Binary}

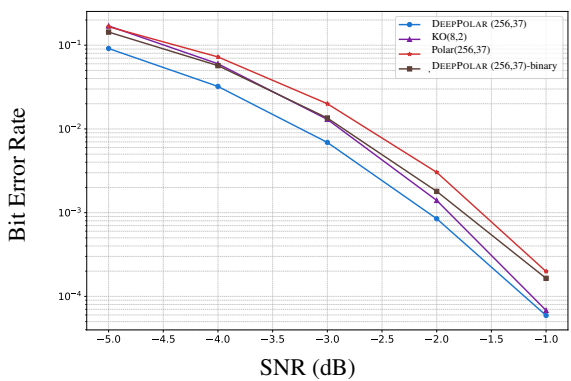
\begin{figure}[t]
\centering{
 \includegraphics[width=\linewidth]{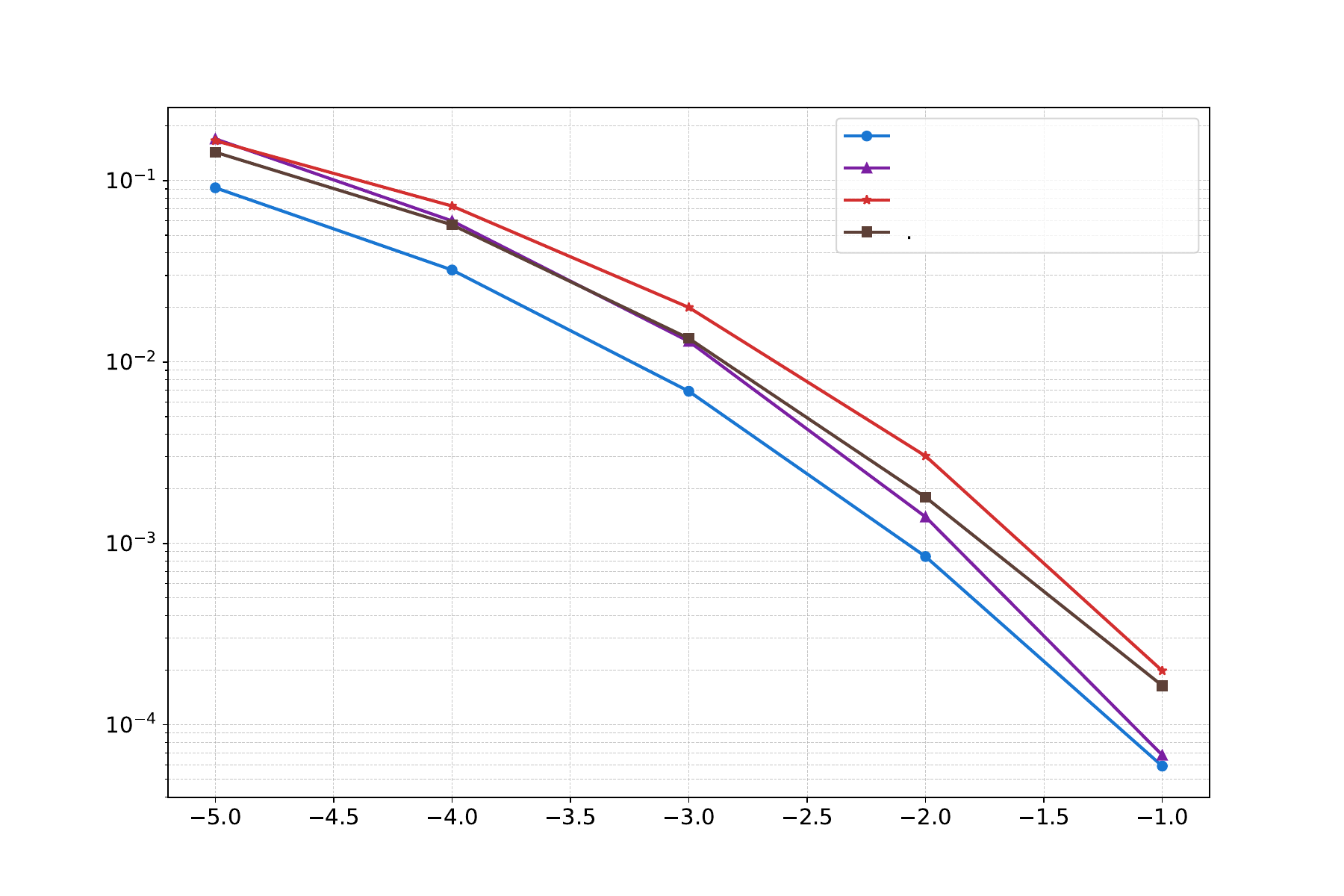}
\put(-75,132){\fontsize{4}{5}\selectfont \dpolar(256,37)}
\put(-75,127){\fontsize{4}{5}\selectfont KO(8,2)}
\put(-75,122){\fontsize{4}{5}\selectfont Polar(256,37)}
\put(-75,116){\fontsize{4}{5}\selectfont \dpolar(256,37)-binary}
\put(-140,0){\footnotesize SNR (dB)}
\put(-235,70){\rotatebox[origin=t]{90}{\footnotesize Bit Error Rate}}
\caption{\dpolar-binary(256,37,$\ell$=16) approaches the performance of KO codes}
\label{fig:binary}
}
\vspace{-0.2in}
\end{figure}

Compared to canonical coding schemes, a real-valued coding scheme like \dpolar offers distinct advantages by integrating modulation and coding schemes. However, practical systems may have hard symbol-level power constraints, necessitating code bits to be integers $x_i \in \{-1,1\}$. Hence, it is beneficial to have the ability to learn a binary code while maintaining the structure of \dpolar.

To achieve this, a trained \dpolar model is fine-tuned with a Straight Through Estimator (STE)~\cite{hubara2016binarized} and a binarization module, resulting in a binary version, \dpolar-binary, akin to the methodology in \cite{jiang2019turbo}. \prettyref{fig:binary} highlights that there is a loss incurred with respect to \dpolar. This underscores the contribution of joint coding and modulation to \dpolar's performance. Nevertheless, \dpolar-binary not only surpasses the performance of the traditional Polar code but also closely matches that of KO codes. Additionally, the distance profile of \dpolar-binary (\prettyref{fig:distance}) indicates it preserves Gaussian codebook-like distance properties, reinforcing the potential of non-linear binary codes derived from polarization-inducing structures. This exploration opens avenues for future research, particularly in interpreting these codes to develop novel, efficient encoders and decoders leveraging non-linear polarization-based methods.



\textbf{Effect of curriculum.}
Ensuring reliable performance of neural codes at high SNR levels is a significant challenge, primarily due to the sparse occurrence of error events in this regime. This challenge often leads to a phenomenon known as an \textit{error floor}, where the error rate improvement of a code stagnates beyond a certain SNR \cite{jamali2021productae}. This issue is not exclusive to neural codes, as it's also encountered in classical codes like LDPC and Turbo codes. While training hyperparameters such as SNR scheduling and batch sizes affect high SNR performance - the structure of the code also plays a part. Notably, algebraic codes like polar codes do not suffer from error floors. Driven by this intuition, our curriculum design (\prettyref{sec:curriculum}) reinforces the Plotkin code structure, leading to better generalization. This strategy has been effective, as highlighted in \prettyref{fig:kernel_size_curriculum}, where implementing a curriculum markedly enhances code performance at higher SNRs.

\textbf{DeepPolar decoder for Polar codes}

\begin{figure}[t]
\centering{
 \includegraphics[width=\linewidth]{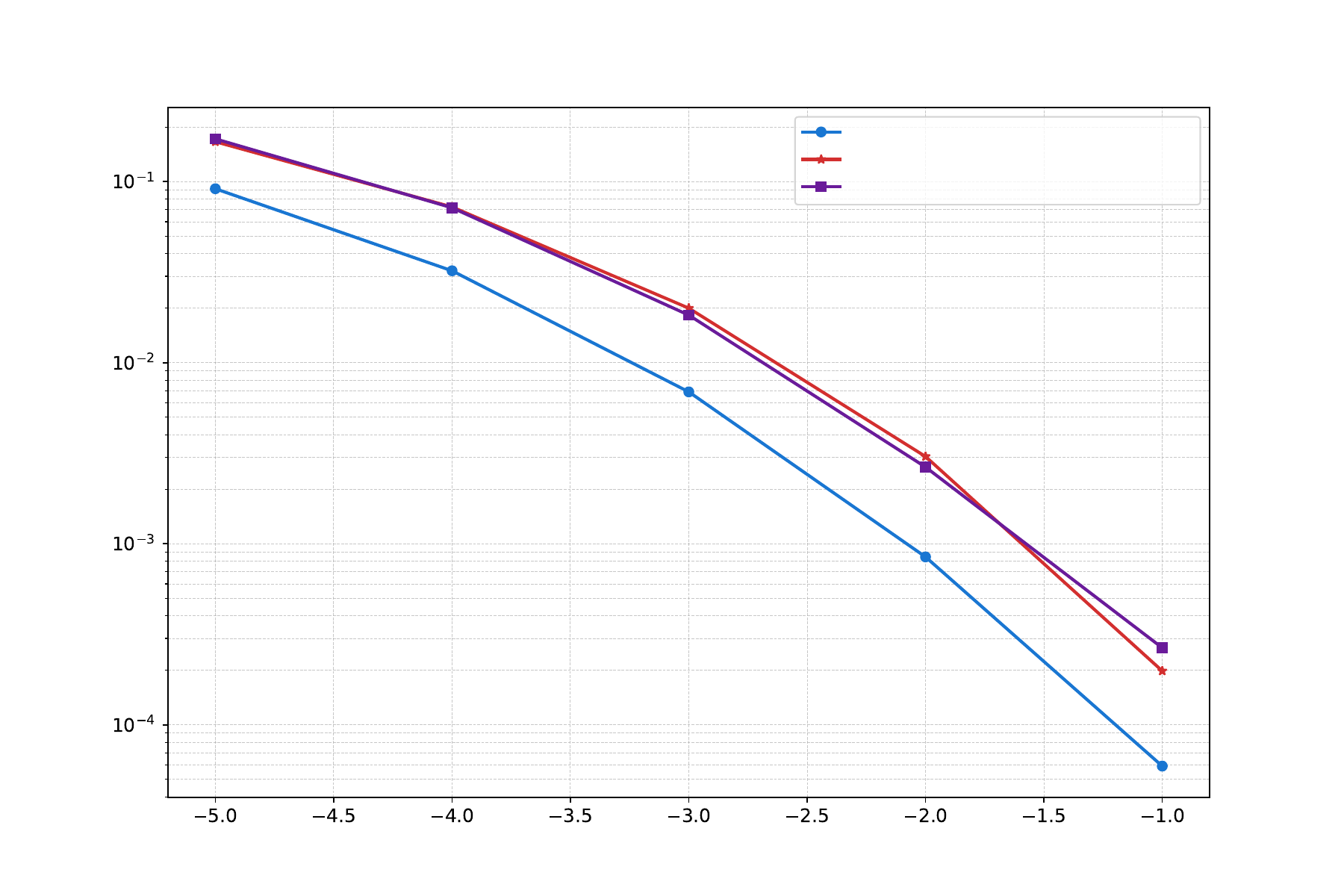}
\put(-85,132){\fontsize{4}{5}\selectfont \dpolar encoder+decoder}
\put(-85,127){\fontsize{4}{5}\selectfont Polar encoder+decoder}
\put(-85,122){\fontsize{4}{5}\selectfont Polar encoder+ \dpolar decoder}
\put(-140,0){\footnotesize SNR (dB)}
\put(-235,70){\rotatebox[origin=t]{90}{\footnotesize Bit Error Rate}}
\caption{For decoding Polar(256,37) codes, \dpolar decoder matches the SC decoder.}
\label{fig:learned_decoder}
}
\end{figure}
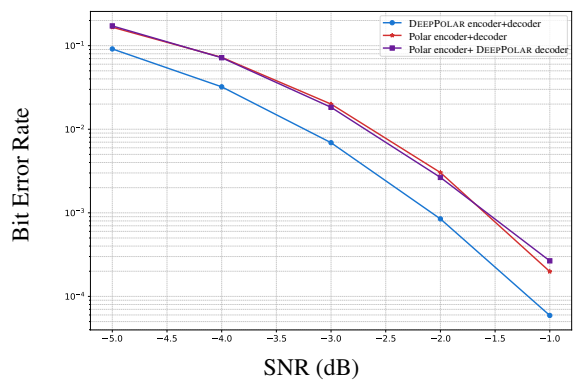

\dpolar codes are a generalization of Polar codes, and inherits its encoding structure (Plotkin tree, frozen set), and the sequential decoding paradigm, and achieves notable gains in BER. To pinpoint the efficacy of the learned code, we fix the Polar encoder and train the \dpolar decoder to predict the transmitted message. 
\prettyref{fig:learned_decoder} demonstrates that the \dpolar decoder matches the performance of SC decoding to decode Polar codes. This is expected, owing to the SC-like parameterization of the \dpolar decoder. The gains in BER, thus, can largely be attributed to the encoder’s non-linear, binary-to-real encoding scheme.

\subsection{Computational Complexity}
In our study, we introduce a novel non-linear generalization of the Polar code's encoding and decoding structures and achieve substantial improvements in reliability. Another essential aspect of evaluating an algorithm is the \textit{computational complexity}, which has a direct impact on power consumption. Polar codes are favored due to their relatively low-complexity encoding and decoding algorithms. However, in practice, CRC-aided list decoders are used, which substantially increases the decoding complexity. 
\dpolar is a non-linear generalization of Polar codes via neural-network-based kernels on the Plotkin tree. The \dpolar SC-decoder is a kernel-wise sequential decoder inspired by the SC algorithm. For the case of \dpolar$(n=256,k=37,\ell=16)$, the encoder and decoder have 0.1M and 1.6M parameters, respectively, which we did not optimize in this project.
Our findings serve as constructive proof that non-linear polar codes with large kernels can outperform the state-of-the-art. Further, we demonstrate the viability of effective binary codes with a large-kernel Polar structure, complete with efficient decoders. This opens avenues for further examination of these codebooks to inform the development of low-complexity solutions. Notably, the complexity of neural codes can be reduced significantly without performance degradation by complexity-aware architecture design (eg, TinyKO \cite{Makkuva2021}). Our preliminary experiments with \dpolar-parallel (\prettyref{app:ell_parallel}), which reduces the decoder parameter count by $8\times$, have already demonstrated \textit{improved} performance at lower complexity. Further, several techniques are used in practice, like distillation, quantization, and pruning, among others, to reduce the computational overhead of a neural algorithm. This is important and interesting future work beyond the scope of this paper.


\section{Related work}

The application of machine learning to channel coding has been an active area of research in recent years. The majority of existing works focus on decoding existing codes, aiming to achieve better reliability and robustness, and in some cases, lower decoding complexity \cite{kim2018communication, nachmani2016learning, dorner2017deep, vasic2018learning, nachmani2019hyper, jiang2019deepturbo, chen2021cyclically, jamali2021reed,  choukroun2022denoising, choukroun2022error, hebbar2022tinyturbo, aharoni2023data}. In the context of polar codes, several neural decoders have been proposed \cite{cammerer2017combining, cammerer2017scaling, doan2018neural,  hebbar2023crisp}. In another line of work, \cite{ebada2019deep, liao2021construction, li2021learning, miloslavskaya2023design,anki2024nested} use deep learning to identify optimal polar code frozen positions without altering the design of encoder and decoder.

In contrast, jointly learning both encoders and decoders is a more challenging problem, and very few works in the literature exist. A common theme is the incorporation of principled coding-theoretic encoding and decoding structures. TurboAE \cite{jiang2019turbo}, and follow-up works \cite{jiang2020feedback, jiang2020joint, chahine2021deepic, saber2022list, chahine2022turbo, wang2023rate}, use sequential encoding and decoding along with interleaving of input bits, inspired by Turbo codes \cite{Berrou1993}. KO codes \cite{Makkuva2021} generalize Reed-Muller encoding and Dumer decoding by replacing selected components in the Plotkin tree with neural networks. ProductAE \cite{jamali2021productae} generalizes two-dimensional product codes to scale neural codes to larger block lengths. In a similar vein, our work generalizes the coding structures of large-kernel Polar codes by using non-linear kernels parameterized by NNs.
\looseness=-1

Deep learning-based schemes for channels with feedback is another active area of research \cite{kim2018deepcode, safavi2021deep, chahine2022inventing, ozfatura2022all, ozfatura2023feedback, ozfatura2023not, kim2023robust, ankireddy2024lightcode}
Furthermore, breaking the conventional wisdom that neural codes are hard to interpret, 
\cite{Devroye2023interpret} derives a closed-form approximation of binarized TurboAE codes via mixed integer linear programming and other techniques. The task of analytically approximating and understanding binarized \dpolar codes remains a promising subject for future research. 


\section{Conclusion}
In this work, we introduce \dpolar codes, a new class of non-linear generalizations of large-kernel Polar codes. \dpolar codes generalize to various rates and blocklengths, and outperform the current state-of-the-art neural codes in BER. The neural architecture mirrors the Polar encoding and decoding structures, which along with a curriculum training methodology is key to improve generalization to unseen messages.


\section*{Acknowledgement} 
This research is supported in part by NSF CCF 2312753, ONR N00014-21-1-2388, NSF CNS 2112471, ARO W911NF2310062, ONR N00014-21-1-2379, NSF CNS-2008824, a gift from Intel, and Samsung Research America through 6G$@$UT center within the Wireless Networking and Communications Group (WNCG) at the University of Texas at Austin. We thank P. Trifonov for his insightful comments.

\section*{Impact Statement} This paper presents work whose goal is to advance the field of channel coding using machine learning. There are many potential societal consequences of our work, none of which we feel must be specifically highlighted here.

\newpage

\bibliography{references}
\bibliographystyle{icml2024}

\newpage
\appendix
\onecolumn
\section{Successive Cancellation decoder}\label{app:sc_decoder}

 Here we look at the Successive Cancellation Decoding algorithm, provided in \cite{arikan2009channel}. As the name suggests, the SC decoding algorithm decodes the bits sequentially, starting with $u_0$. A frozen bit node is always decoded as $0$. And while decoding the $u_i$, the available bits $u_0$ to $u_0^{i-1}$ which are represented by the vector $u_{i-1}$ are used according to decode $u_i$ according to the following rule:
 \begin{equation}
      \hat{u_i} = \left\{
  \begin{array}{@{}ll@{}}
    0, & \text{if}\ i \in \mathcal{I}\ \text{and}\ Pr[y,\hat{u_0}^{i-1}|u_i=0] \geq Pr[y,\hat{u_0^{i-1}|u_i=1}]\\
    1, & \text{if}\ i \in \mathcal{I}\ \text{and}\ Pr[y,\hat{u_0}^{i-1}|u_i=0] \leq Pr[y,\hat{u_0^{i-1}|u_i=1}]\\
    0, & \text{if}\ i \in \mathcal{F}
  \end{array}\right.
 \end{equation}

 where $A$ is the set of information positions.
 
 The probability calculations are computationally easier and less prone to round-off errors in the log domain. Hence, we consider LLRs instead of probabilities to avoid numerical overflows. LLR for $t^{th}$ bit is defined as:
 $$
 L^{(i)}(y,\hat{u_0}^{i-1}) = log{\dfrac{Pr[y,\hat{u_0}^{i-1}|u_i=0]}{Pr[y,\hat{u_0^{i-1}|u_i=1}]}}
 $$
 Hence, the decision rule changes as follows:
 \begin{equation}
       \hat{u_i} = \left\{
  \begin{array}{@{}ll@{}}
    0, & \text{if}\ i \in \mathcal{I}\ \text{and}\ L^{(i)}(y,\hat{u_0}^{i-1}) \geq 0\\
    1, & \text{if}\ i \in \mathcal{I}\ \text{and}\ L^{(i)}(y,\hat{u_0}^{i-1}) \leq 0\\
    0, & \text{if}\ i \in \mathcal{F}
  \end{array}\right.
 \end{equation}
 
 The binary tree structure of a polar code can be exploited to simplify the successive cancellation decoding. The binary tree has $n = \log_2 N + 1$ stages with numbering from $s = {0,\dots,n}$. Each stage $s$ contains $2^s$ nodes with each node corresponding to $2^{n-s}$ bits.

 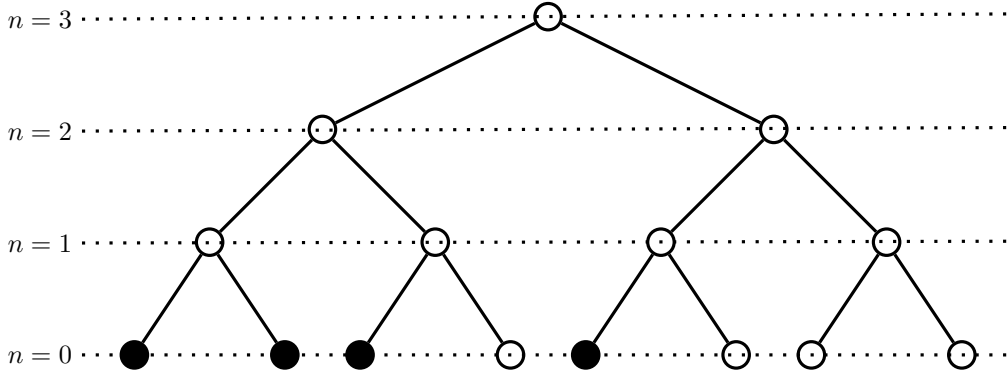
\begin{figure}
 \centering
\tikzstyle{vertex}=[draw,fill=black!0,circle,minimum size=10pt,inner sep=0pt]
 \tikzstyle{vertex_info}=[draw,fill=black!0,circle,minimum size=10pt,inner sep=0pt]
 \tikzstyle{vertex_frozen}=[draw,fill=black!100,circle,minimum size=10pt,inner sep=0pt]
\begin{tikzpicture}[very thick,level/.style={sibling distance=60mm/#1},
empty/.style={rectangle, draw=white!100, fill=white!5, thick, minimum width=0cm, minimum height=0cm}]
\node [vertex]  (r0) {}
  child {
    node [vertex]  (r1) {}
    child {
      node [vertex]  (r2) {}
      child {
        node [vertex_frozen]  (r3) {}
      }
      child {node [vertex_frozen] {}}
    }
    child {
      node [vertex] {}
      child {node [vertex_frozen] {}}
      child {node [vertex] {}}
    }
  }
  child {
    node [vertex] {}
    child {
      node [vertex] {}
      child {node [vertex_frozen] {}}
      child {node [vertex] {}}
    }
    child {
      node [vertex] {}
      child {node [vertex] {}}
      child {node [vertex] (r3_last) {}}
    }
  };
  
\node[empty] (r3_st) [left = 0.5cm of r3] {$n = 0$};
\node[empty] (r3_en) [right = 0.5cm of r3_last] {};

\node[empty] (r2_st) [above = 1cm of r3_st] {$n = 1$};
\node[empty] (r2_en) [above = 1.25cm of r3_en] {};

\node[empty] (r1_st) [above = 1cm of r2_st] {$n = 2$};
\node[empty] (r1_en) [above = 1.25cm of r2_en] {};

\node[empty] (r0_st) [above = 1cm of r1_st] {$n = 3$};
\node[empty] (r0_en) [above = 1.25cm of r1_en] {};

\draw[loosely dotted] (r3_st.east) -- (r3_en.west);
\draw[loosely dotted] (r2_st.east) -- (r2_en.west);
\draw[loosely dotted] (r1_st.east) -- (r1_en.west);
\draw[loosely dotted] (r0_st.east) -- (r0_en.west);

\end{tikzpicture}
\caption{Binary tree structure of (8,4) polar code with stages indicated. The frozen bits (in black) are set to 0. Operations at each node are detailed in ~\prettyref{fig:sc_dec_rule}.}
\label{binary_tree_level}
\end{figure}

 In order, traversal of the tree is done to perform the successive cancellation decoding. At each node, messages are passed as shown:
 
  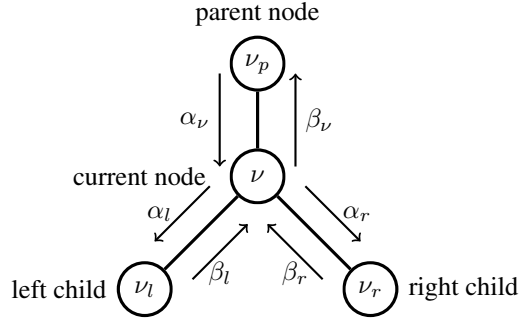
\begin{figure}
 \centering
\tikzstyle{vertex}=[draw,fill=black!0,circle,minimum size=20pt,inner sep=0pt]
 \tikzstyle{vertex_info}=[draw,fill=black!0,circle,minimum size=20pt,inner sep=0pt]
 \tikzstyle{vertex_frozen}=[draw,fill=black!30,circle,minimum size=20pt,inner sep=0pt]
\begin{tikzpicture}[very thick,level/.style={sibling distance=60mm/#1},
empty/.style={rectangle, draw=white!100, fill=white!5, thick, minimum width=0cm, minimum height=0cm}]
\node [vertex] (parent) [label=above:parent node] {$\nu_p$}
  child {
    node [vertex] (current) [label=left:current node .] {$\nu$}
    child {
      node [vertex] (left_child) [label=left:left child] {$\nu_l$}
    }
    child {
      node [vertex] (right_child)  [label=right:right child] {$\nu_r$}
    }
  };
\node[empty] (lp1) [left = 0cm of parent] {};
\node[empty] (lc1) [left = 0cm of current] {};

\node[empty] (lp2) [right = 0cm of parent] {};
\node[empty] (lc2) [right = 0cm of current] {};
\node[empty] (lc3) [below = 0cm of current] {};

\node[empty] (ll1) [above = 0cm of left_child] {};
\node[empty] (ll2) [right = 0cm of left_child] {};

\node[empty] (lr1) [above = 0cm of right_child] {};
\node[empty] (lr2) [left = 0cm of right_child] {};

\path[->,draw,thick] (lp1) edge node[left] {$\alpha_\nu$} (lc1);
\path[->,draw,thick] (lc2) edge node[right] {$\beta_\nu$} (lp2);
\path[->,draw,thick] (lc1) edge node[left] {$\alpha_l$} (ll1);
\path[->,draw,thick] (ll2) edge node[below] {$\beta_l$} (lc3);
\path[->,draw,thick] (lc2) edge node[right] {$\alpha_r$} (lr1);
\path[->,draw,thick] (lr2) edge node[below] {$\beta_r$} (lc3);

\end{tikzpicture}
\caption{SC decoding update rules for each node}
 \end{figure}\label{fig:sc_dec_rule}

 Each node passes LLR corresponding LLR values, namely $\alpha$, to the child nodes and sends the estimated hard bits at the sage, namely $\beta$, o the parent node. The left and right messages, $\alpha_i^l$ and $\alpha_i^r$ are calculated as:

 \begin{equation}
     \alpha_i^l = \ln\bigg(\dfrac{1 + e^{\alpha_i + \alpha_{i+2^{n-s-1}}}}{e^{\alpha_i} + e^{\alpha_{i+2^{n-s-1}}}}\bigg)\\
 \end{equation}
 
 \begin{equation}
      \alpha_i^r = \alpha_{i+2^{n-s-1}} + (1-2\beta_i^l)\alpha_i
 \end{equation}
 
 We define two functions to perform these operations, namely $f$ and $g$, defined as:

\begin{equation} \label{eq1}
\begin{split}
f(\alpha_1,\alpha_2) & = \ln\bigg(\dfrac{1 + e^{\alpha_1 + \alpha_2}}{e^{\alpha_1} + e^{\alpha_2}}\bigg)\\
 g(\beta,\alpha_1,\alpha_2) & = (1-2\beta)\alpha_1 + \alpha_2
\end{split}
\end{equation}

But the $f$ function is computationally expensive, and hence, we approximate it to a hardware-friendly version using min-sum approximation as follows:

\begin{equation}
    f_{minsum}(\alpha_1,\alpha_2) = \textit{sign}(\alpha_1)\textit{sign}(\alpha_2)\textit{min}(|\alpha_1|,|\alpha_2|)
\end{equation}
 
 where \textit{sign} gives the sign of input and \textit{min} gives the minimum of the two inputs.

The algorithm starts from the root node of the tree, which is level $n+1$, and traverses to the leaf node, which is level $0$. For each node, the following set of operations occurs.
\begin{enumerate}
    \item If the current node has a left child that was not visited, calculate $\alpha_l$ and move to the left child.
    \item If the current node has a right child that was not visited, calculate $\alpha_r$ and move to the right child.
    \item If both the messages from child nodes are available, calculate $\beta$ and move to the parent node.
\end{enumerate}

Once the leaf node is reached, decisions are made based on the sign of the corresponding LLR using the binary quantizer function $h$ as:
\begin{equation}
    \beta_\nu = h(\alpha_nu)
\end{equation}

where $h$ is defined as:
 \begin{equation*}
      h(\alpha) = \left\{
  \begin{array}{@{}ll@{}}
    0, & \text{if}\ \alpha \geq 0\\
    1, & \text{else}\\
  \end{array}\right.
 \end{equation*}

\section{Polar codes}

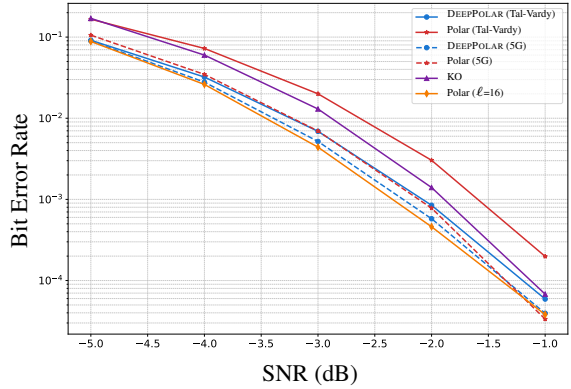
\begin{figure}[h]
\begin{center}
\includegraphics[width=0.5\columnwidth]{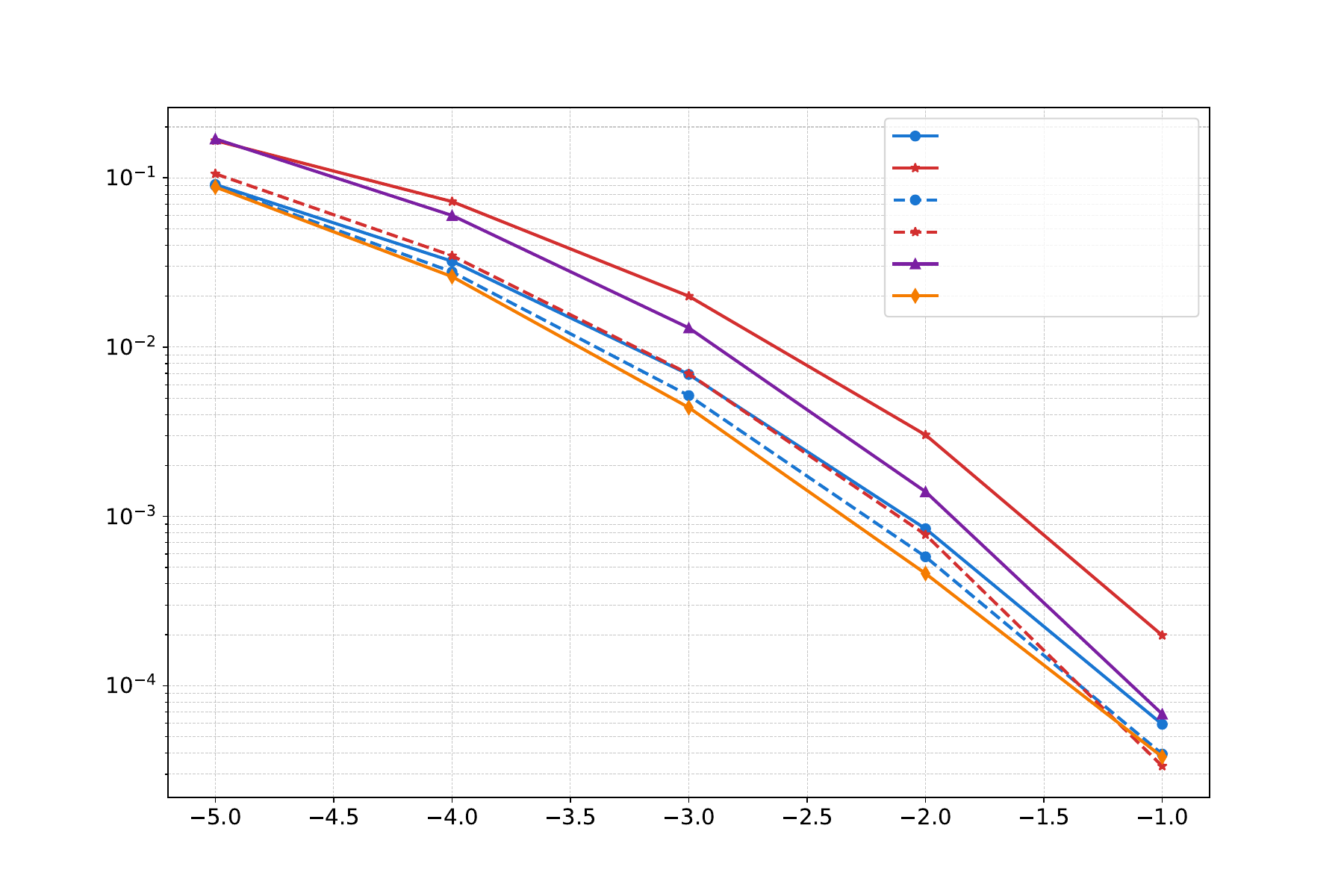}
\put(-70,138){\fontsize{4}{5}\selectfont \dpolar (Tal-Vardy)}
\put(-70,132){\fontsize{4}{5}\selectfont Polar (Tal-Vardy)}
\put(-70,126){\fontsize{4}{5}\selectfont \dpolar (5G)}
\put(-70,120){\fontsize{4}{5}\selectfont Polar (5G)}
\put(-70,114){\fontsize{4}{5}\selectfont KO}
\put(-70,108){\fontsize{4}{5}\selectfont Polar ($\ell$=16)}
\put(-140,0){\footnotesize SNR (dB)}
\put(-235,70){\rotatebox[origin=t]{90}{\footnotesize Bit Error Rate}}
\caption{$n$=256,$k$=37 : \textcolor{black}{The choice of frozen positions impacts the performance of \dpolar. While using 5G frozen positions improves \dpolar, its gains over the 5G-Polar code are modest. \dpolar is a preliminary effort towards an automated search for large-kernel codes; it matches the best performing Polar code with $\ell=16$ \cite{trofimiuk2021window}}}
\label{fig:large_5g}
\end{center}
\end{figure}

\subsection{5G Polar codes}\label{app:5g}
\textcolor{black}{Polar codes are conceptualized by channel polarization, i.e., the recursive application of the kernel results in bit-channels that are either noiseless or very noisy. However, at the practical blocklengths that we consider in this paper, there exist bit channels that are partially polarized. Consequently, the selection of frozen positions 
affects the reliability of polar codes.
Although \dpolar does not exhibit a clear polarization effect—due to its operation in the real field and its non-recursive kernel application—the choice of frozen positions still significantly impacts the effectiveness of the learned code.}

\textcolor{black}{In the paper, we consider polar codes constructed via the Tal-Vardy method as our baseline. This choice is justified by the fact that \dpolar utilizes the Tal-Vardy rate profile, ensuring a fair comparison.} 
\textcolor{black}{ However, we realize that the universal reliability sequence specified in the 5G standards can be used to find a better frozen set. \prettyref{fig:large_5g} depicts that \dpolar trained using the 5G frozen positions shows non-trivial improvement. However, the performance gains compared to the 5G-Polar code are modest.} \textcolor{black}{Exploring joint optimization of rate-matching strategies (such as freezing or precoding) along with training \dpolar presents an intriguing avenue for future research.}

\subsection{Large kernel Polar codes}\label{app:large_polar}

The SC decoding operation for larger kernel Polar codes ($\ell > 2$) is similar to the original polar codes. Iteratively, the decoding equation can be expressed as
\begin{equation*}
    W_m^{(i)}((u_0^{(i-1)},u_i)|y_0^{(n-1)}) = \frac{ W_m^{(i)}(y_0^{n-1},u_0^{i-1}|u_i) }{2W(y_0^{n-1})},
\end{equation*}
which is the probability for path $u_0^i$ given channel output $y_0^{n-1}$.

\textcolor{black}{While these operations are similar to conventional polar codes, the complexity of iteratively computing the probabilities increases exponentially with respect to $\ell$ as $O(2^{\ell})$, making it prohibitively expensive for very large kernels. However, recent work has shown methods towards efficient decoding of a class of large-kernel polar codes \cite{ abbasi2020large, gupta2021polar, trofimiuk2021window}.} 

\textcolor{black}{\dpolar generalizes the conventional polar codes, and makes two notable modifications. First, it expands the kernel size to $\sqrt{n}$. Second, it introduces neural networks to parameterize the encoding and decoding functions for each kernel, which are then learned through training. The exhaustive search across large kernel spaces to improve upon Arikan's original construction is infeasible. Nonetheless, recent advancements have been made by limiting this space and employing more efficient decoding methods \cite{trifonov2023design}. \dpolar represents an initial attempt to automate the search for effective large-kernel polar codes.}

\textcolor{black}{Here, we consider an optimized construction of a kernel of size $\ell = 16$ ~\cite{trofimiuk2021window}, with SC decoding and compare it with \dpolar code of $\ell=16$. \prettyref{fig:large_5g} shows that \dpolar marginally falls short of this optimized construction. An exciting future direction would be to directly augment these codes using our methodology, akin to how KO codes augmented the Reed Muller code structure.}

\section{\dpolar Decoders}
\subsection{\dpolar-SC Decoder} \label{app:ell_decoder}

\begin{figure}[h]
\begin{center}
\includegraphics[width=0.6\columnwidth, trim={0cm, 0cm, 0cm, 6cm}, clip]{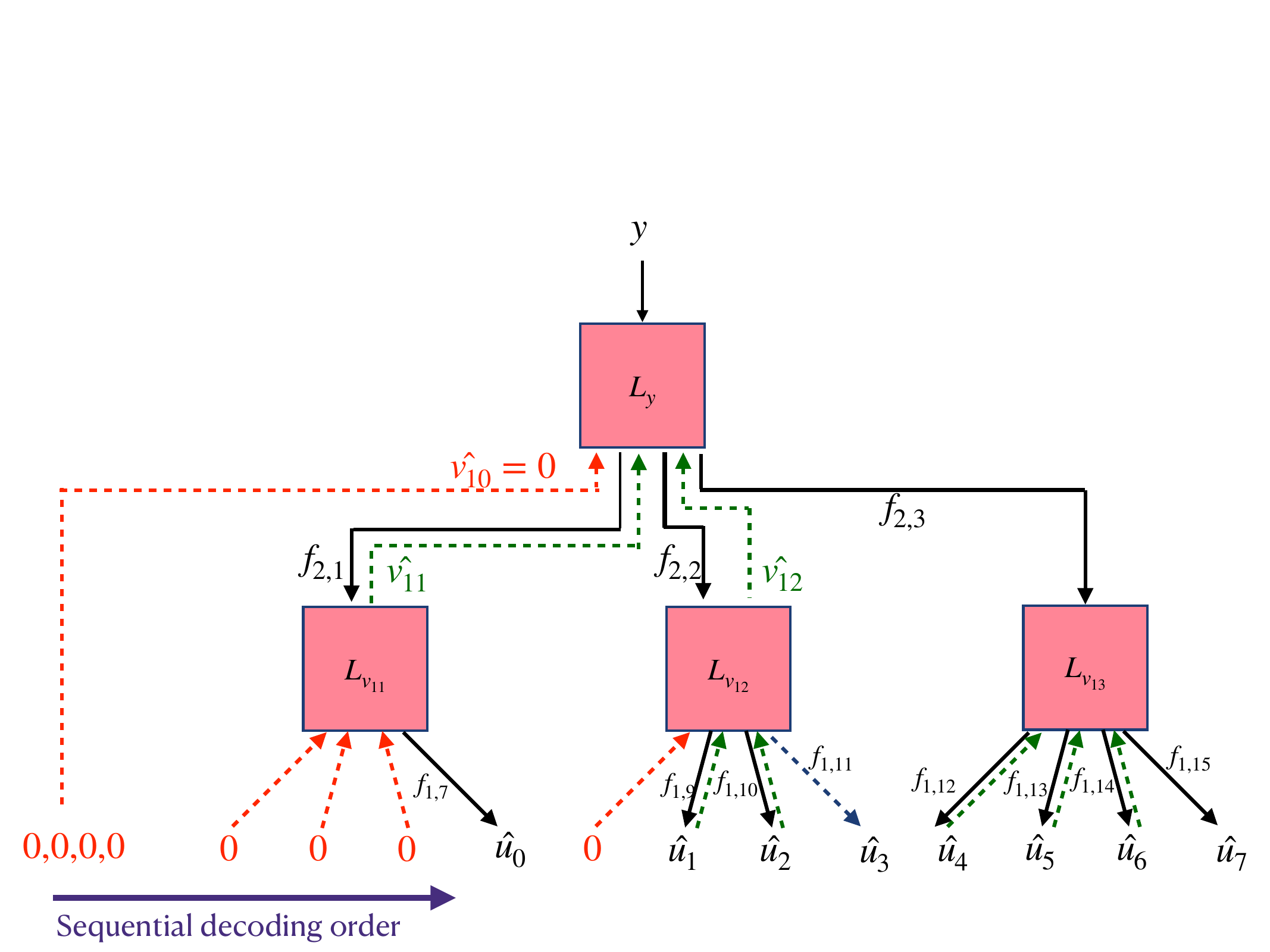}
\caption{\dpolar($n$=16,$k$=8,$\ell$=4) : \dpolar-SC decoder decodes each bit sequentially - each edge operation in the decoding tree is parameterized by a neural network $f_{d,b}$ }
\label{fig:large_polar_dec}
\end{center}
\end{figure}

Decoding of \dpolar for large kernels is depicted in ~\prettyref{fig:large_polar_dec} for \dpolar(16,8,$\ell=4$). The frozen positions are not decoded and are directly set to 0, including the full of the first kernel. Beginning with the first information position, $\hat{u}_0$ is estimated based on the received values $\by$ using sub-kernel $f_{1,7}$. Now, the decoding of the two kernels is complete and sent to the parent node to begin the decoding of kernel 3. Using the received values $\by$ and $\hat{u}_0$, sub-kernel $f_{1,9}$ estimates the next non-frozen position $\hat{u}_1$. Next, using $\by$ and $\{\hat{u}_0,\hat{u}_1 \}$, sub-kernel $f_{1,10}$ decodes the next non-frozen position $\hat{u}_2$. This process is continued until kernel-3 is decoded completely, after which the decoding of kernel-4 starts in the same manner and continues until all bits are estimated.

To summarise, successive cancellation decoding of \dpolar involves sequentially decoding one kernel of size $\ell$ at a time. Decoding the kernel is again performed in a successive fashion, where each bit is decoded by the corresponding component decoder or sub-kernel. 

\subsection{\dpolar-parallel decoder} \label{app:ell_parallel}

\begin{figure}[h]
\begin{center}
\includegraphics[width=0.6\columnwidth, trim={0cm, 0cm, 0cm, 6cm}, clip]{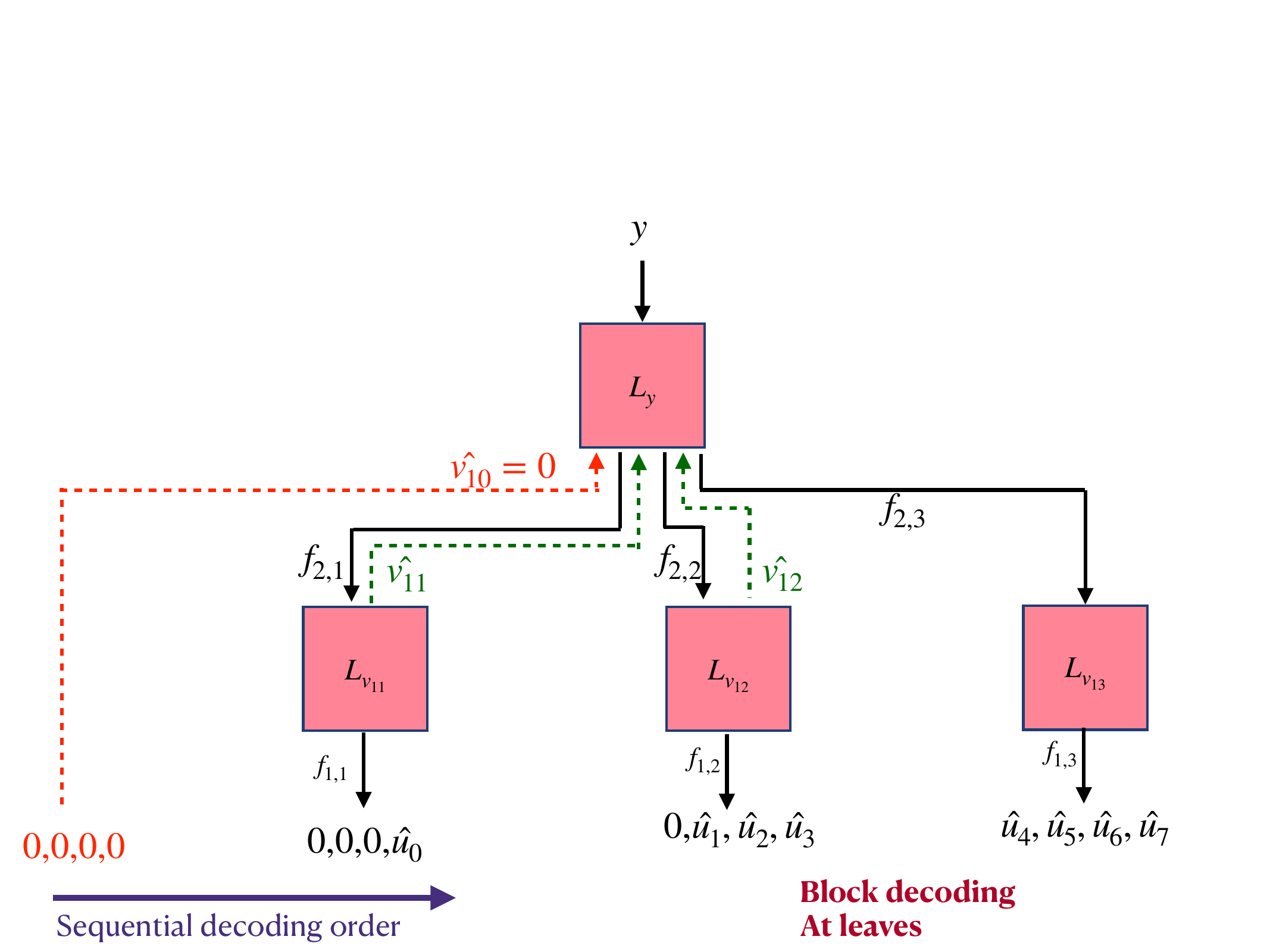}
\caption{\dpolar($n$=16,$k$=8,$\ell$=4) : \dpolar-parallel decoder retains the same decoding structure as \dpolar-SC, but uses a one-shot parallel decoder at the leaves.}
\label{fig:decoder_parallel}
\end{center}
\end{figure}

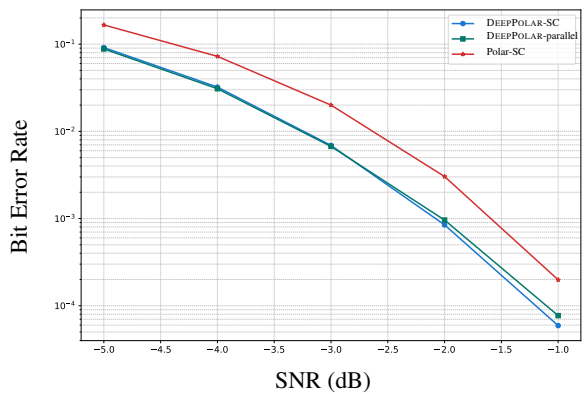
\begin{figure}[ht]
\begin{center}
\includegraphics[width=0.5\columnwidth]{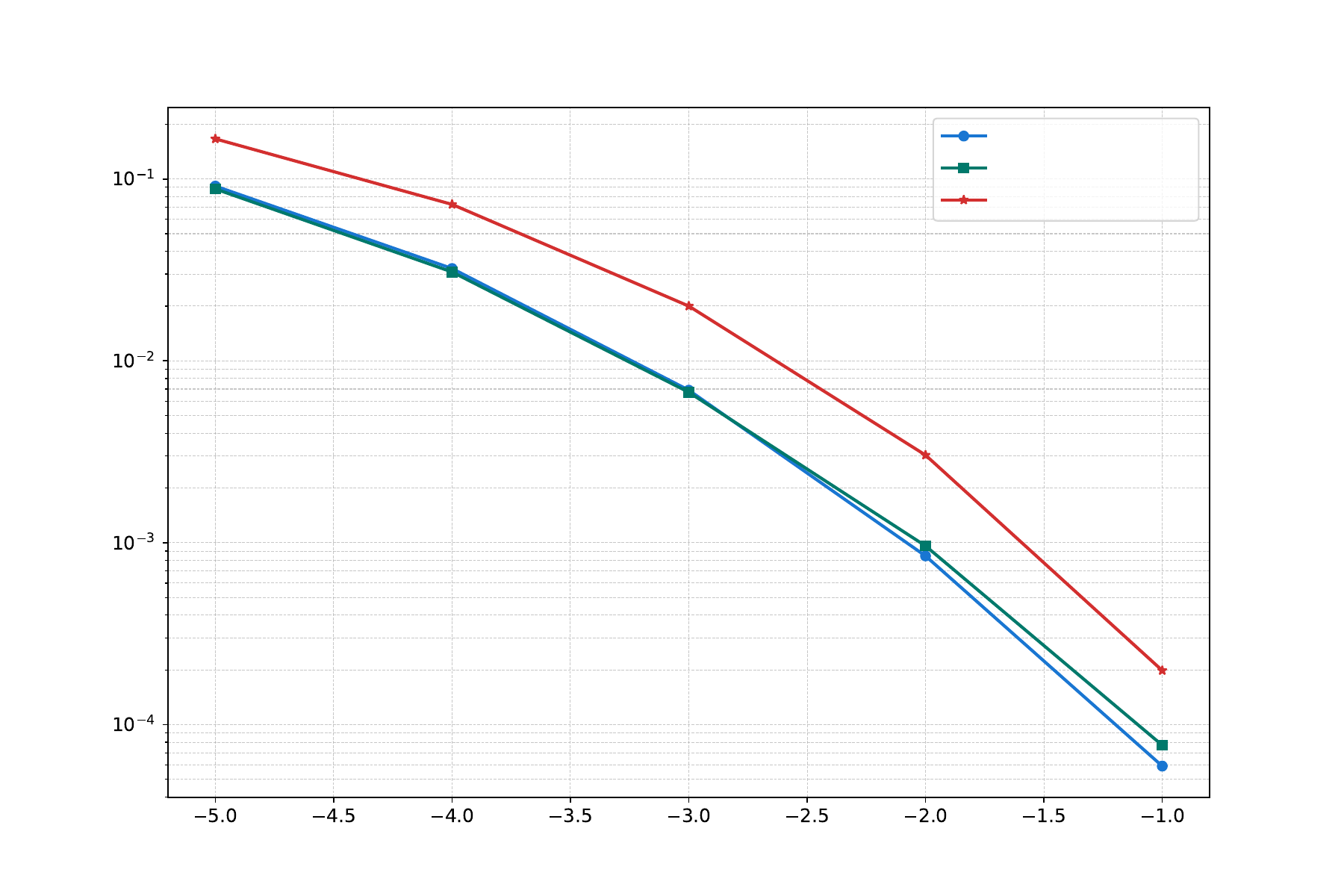}
\put(-60,137){\fontsize{4}{5}\selectfont \dpolar-SC}
\put(-60,132){\fontsize{4}{5}\selectfont \dpolar-parallel}
\put(-60,126){\fontsize{4}{5}\selectfont Polar-SC}
\put(-140,0){\footnotesize SNR (dB)}
\put(-240,70){\rotatebox[origin=t]{90}{\footnotesize Bit Error Rate}}
\caption{$n$=256,$k$=37 : \dpolar-parallel, a lower complexity decoder constructed by replacing the SC-style decoding at the leaves by a bit-MAP-style decoding, achieves better reliability than \dpolar-SC, whilst suffering lower latency } 
\label{fig:parallel}
\end{center}
\end{figure}

A major drawback of SC decoding and its variants like SCL decoders is the latency overhead due to sequential decoding. This drawback is inherently present in the \dpolar-SC decoder. One method to break this latency barrier, whilst reducing the computational complexity of the decoder is to replace the $\ell$ SC-style NN decoders for a kernel at the lowest depth, by a single 3-layer FCNN, with hidden size 64, that decodes $\ell$ bits in one shot, as depicted in \prettyref{fig:decoder_parallel}. We refer to this as \dpolar-parallel. This resulted in an immediate reduction of 8x in parameter count compared to \dpolar-SC while achieving the same reliability, as highlighted in~\prettyref{fig:parallel}.

These preliminary results hint at the existence of low-complexity architectures that provide comparable performance at a much lower complexity. This is a promising direction of future exploration.

\section{Robustness to non-AWGN channels}\label{app:robustness}
Traditionally, the canonical setting for the design and evaluation of codes has predominantly focused on AWGN channels due to the relative simplicity of closed-form performance analysis. However, practical communication channels often deviate significantly from the AWGN model. Ideally, both the code and decoder should be robust to various channels or adaptable to the encountered channel conditions. In this section, we demonstrate that \dpolar codes, which are trained on AWGN channels, exhibit \textit{robustness} when tested on non-AWGN channels \textit{without retraining}. Further, in cases where the target channel model significantly diverges from AWGN noise, data-driven codes like \dpolar can achieve substantial performance improvements through fine-tuning with a small number of steps on the target channel.

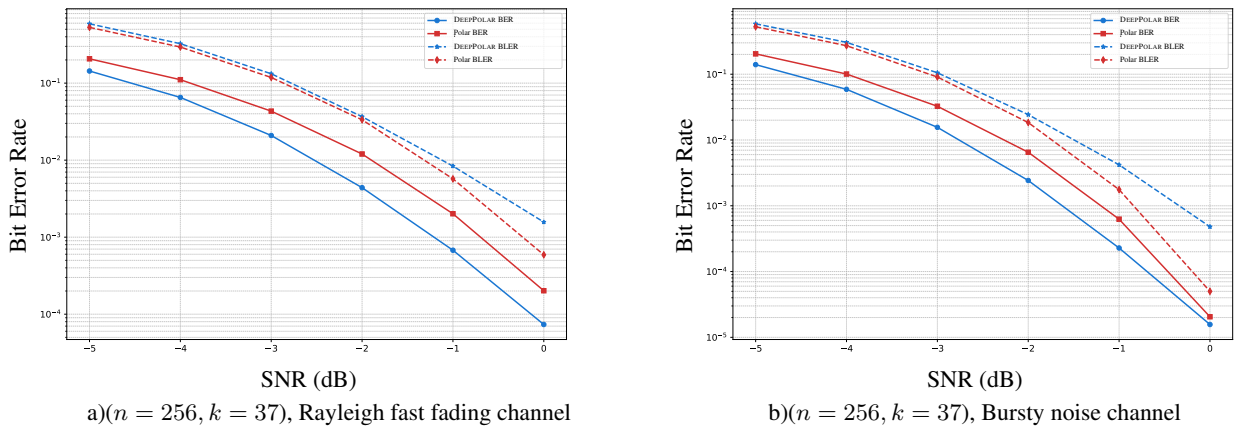
\begin{figure*}[h!]
\captionsetup[subfloat]{labelformat=simple, labelsep=none, listofformat=subsimple}
\renewcommand{\thesubfigure}{\alph{subfigure})}
\centerline{
\subfloat[($n=256,k=37$), Rayleigh fast fading channel]
{  \hspace*{-0.25in}
 \includegraphics[width=0.5\linewidth]{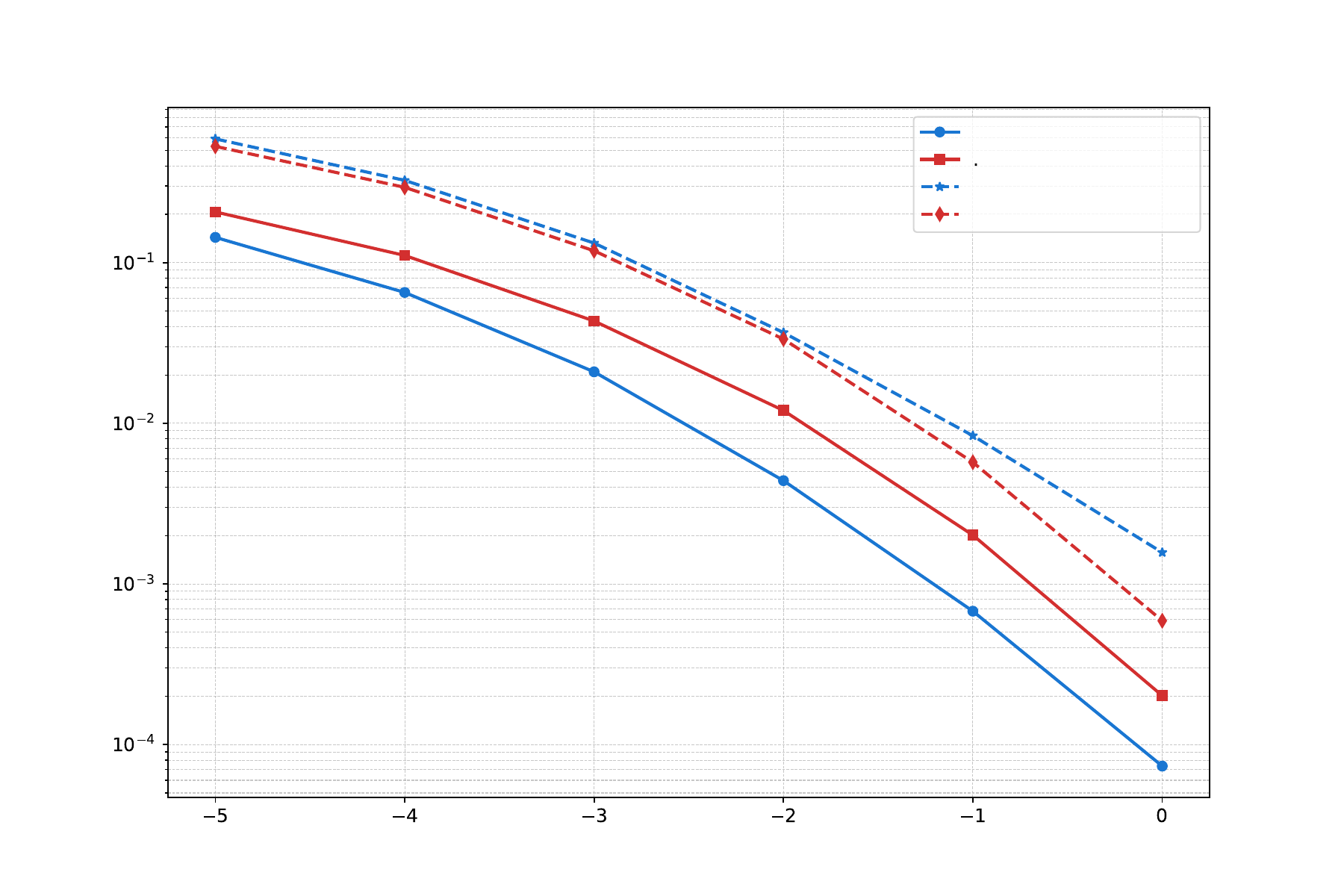} 
    \put(-67,138){\fontsize{3}{5}\selectfont \dpolar BER}
    \put(-67,133){\fontsize{3}{5}\selectfont Polar BER}
    \put(-67,128){\fontsize{3}{5}\selectfont \dpolar BLER}
    \put(-67,123){\fontsize{3}{5}\selectfont Polar BLER}
  \put(-140,0){\footnotesize SNR (dB)}
\put(-235,70){\rotatebox[origin=t]{90}{\footnotesize Bit Error Rate}}
  \label{fig:fading}
 }
\hfill
\subfloat[($n=256,k=37$), Bursty noise channel]
{
 \includegraphics[width=0.5\linewidth]{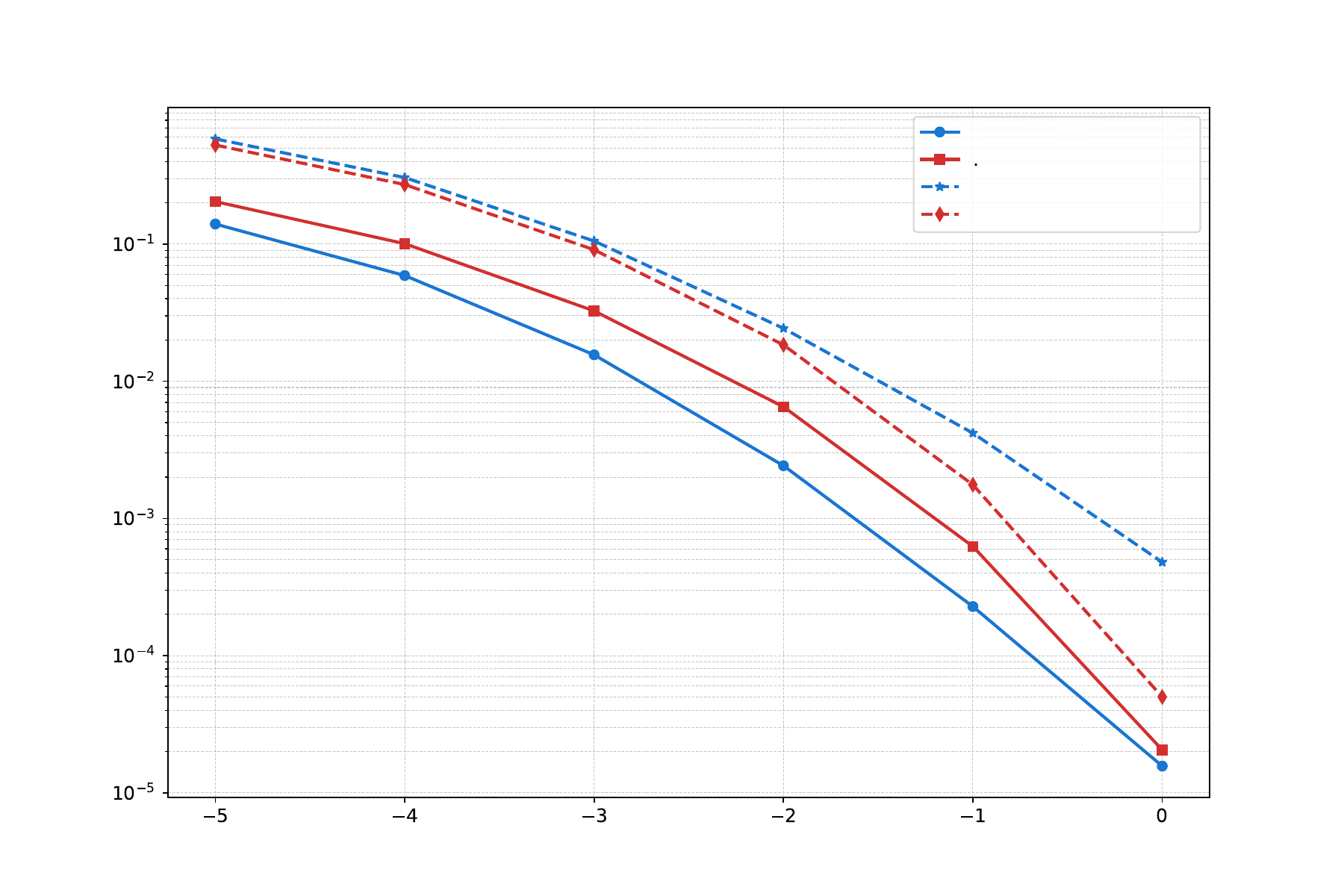}
    \put(-67,138){\fontsize{3}{5}\selectfont \dpolar BER}
    \put(-67,133){\fontsize{3}{5}\selectfont Polar BER}
    \put(-67,128){\fontsize{3}{5}\selectfont \dpolar BLER}
    \put(-67,123){\fontsize{3}{5}\selectfont Polar BLER}
  \put(-140,0){\footnotesize SNR (dB)}
\put(-235,70){\rotatebox[origin=t]{90}{\footnotesize Bit Error Rate}}
  \label{fig:bursty}
}
}
\caption{Robustness to channel variations: \dpolar($256,37,\ell=16$) trained on AWGN channel maintains the gains over $\polar(256,37)$ when tested on Rayliegh fast fading and bursty noise channels.}
\label{fig:robustness}

\end{figure*}
\begin{figure*}[h!]
\captionsetup[subfloat]{labelformat=simple, labelsep=none, listofformat=subsimple}
\renewcommand{\thesubfigure}{\alph{subfigure})}
\centerline{
\subfloat[($n=256,k=37$), Rayleigh fast fading channel]
{  \hspace*{-0.25in}
 \includegraphics[width=0.5\linewidth]{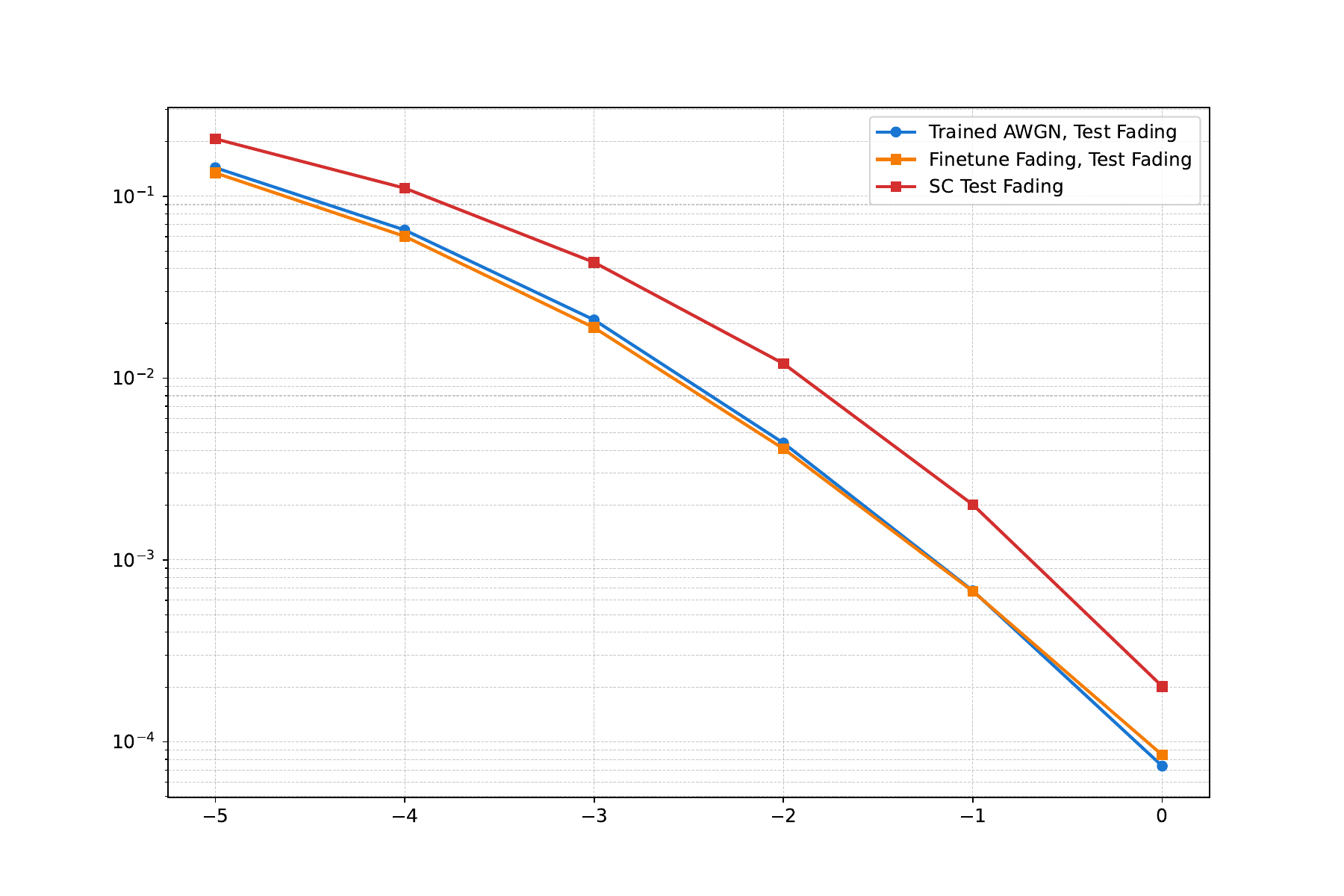} 
    \put(-140,0){\footnotesize SNR (dB)}
\put(-235,70){\rotatebox[origin=t]{90}{\footnotesize Bit Error Rate}}\label{fig:fading_adaptivity}
 }
\hfill
\subfloat[($n=256,k=37$), Bursty noise channel]
{
 \includegraphics[width=0.5\linewidth]{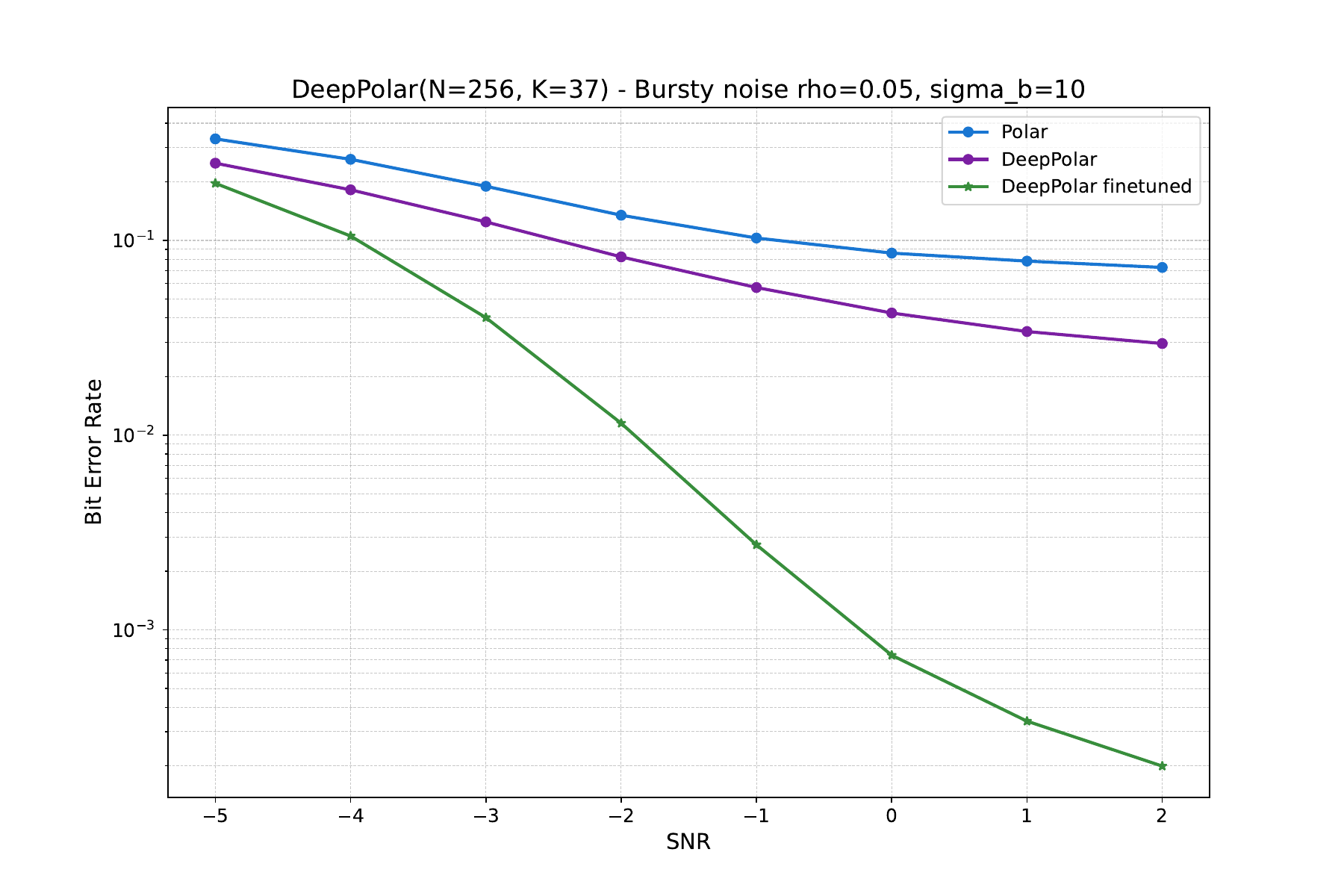}
  \put(-140,0){\footnotesize SNR (dB)}
\put(-235,70){\rotatebox[origin=t]{90}{\footnotesize Bit Error Rate}}
  \label{fig:bursty_adaptivity}
}
}
\caption{Adaptivity: Performance of \dpolar($256,37,\ell=16$) can be improved by adapting it to target channels. On channels such as Rayleigh fast fading, we do not see extra gains over the base decoder trained on AWGN. However, on large bursty noise we see a pronounced improvement}
\label{fig:adaptivity}

\end{figure*}
First, we consider Rayleigh fast-fading channel with AWGN noise. Rayleigh fading is considered a suitable model for signal propagation in tropospheric and ionospheric environments, as well as for the effect of heavily built-up urban areas on radio signals. Mathematically, the channel model can be approximated as 
\begin{equation*}
    y_i = h_i x_i + z_i,
\end{equation*}
where $h \sim \mathcal{N}(0,1)$ is the fading coefficient and $z_i \sim \mathcal{N}(0,\sigma^2)$ is the Gaussian noise. Note that the fading coefficient changes for every symbol in the codeword, making it a much worse channel than the AWGN channel. As seen from ~\prettyref{fig:fading}, the \dpolar code demonstrates a gain of up to 0.3 dB compared to $\polar(256,37)$, demonstrating the robustness of the learned code. 

Next, we consider a bursty noise channel. Bursty noise, also known as popcorn noise, can be described mathematically as
\begin{equation*}
    y_i = x_i + z_i + w_i,
\end{equation*}
where $z_i \sim \mathcal{N}(0,\sigma^2)$ is the Gaussian noise and $w_i \sim \mathcal{N}(0,\sigma^2_b)$ with probability $\rho$ and $w_i = 0$ with probability $1 - \rho$ is the bursty noise. For our experiment, we choose $\rho = 0.1$ and $\sigma_b = \sqrt{2} \sigma$. As seen from ~\prettyref{fig:bursty}, the \dpolar code demonstrates a gain of up to 0.25 dB compared to $\polar(256,37)$ in the presence of bursty noise.

Under larger degradation, for instance a bursty channel with $\rho=0.05$ and burst power $\sigma_b$= $10$, DeepPolar maintains its robustness. Still, the performance can be greatly enhanced by fine-tuning the decoder specifically for the target channel model, as demonstrated in ~\prettyref{fig:bursty_adaptivity}. This adaptivity to unseen channel models is a key advantage of neural codes and decoders, such as DeepPolar, compared to classical codes.

\section{Experimental details} \label{sec:experimental}
The complete source code is provided at: \href{https://www.github.com/hebbarashwin/deeppolar}{https://www.github.com/hebbarashwin/deeppolar}

\subsection{Training algorithm}
We follow a training strategy similar to KO codes~\cite{Makkuva2021}, where the encoder and decoder networks are trained in an alternating optimization, described below. 

\begin{algorithm}
\caption{Training algorithm for \dpolar(256,37,$\ell$=16)}
\label{alg:trainingKO}
\begin{algorithmic}[1]
\STATE Initialize encoder and decoder parameters
\FOR{$E$ epochs}
  \FOR{$T_{\text{dec}}$ steps}
    \STATE Generate $B$ random message vectors $\bu$
    \STATE Simulate AWGN channel with $\text{SNR}_{\text{dec}}$
    \STATE Freeze the encoder $g$ and update the decoder network $f$ to minimize the BCE loss $L(g,f)$ using Adam and learning rate $lr_{\text{dec}}$.
  \ENDFOR
  \FOR{$T_{\text{enc}}$ steps}
    \STATE Generate $B$ random message vectors $\bu$
    \STATE Simulate AWGN channel with $\text{SNR}_{\text{enc}}$
    \STATE Freeze the decoder $f$ and update the encoder network $g$ to minimize the BCE loss $L(g,f)$ using Adam and learning rate $lr_{\text{enc}}$.
  \ENDFOR
\ENDFOR
\end{algorithmic}
\end{algorithm}

\subsection{Training SNR and number of steps}
Choosing the right SNR is critical to avoid local optima during training. Choosing a very low SNR will result in the noise dominating the training data, and as a result, the training may not converge. On the contrary, choosing a very high SNR will not provide sufficient error events for the model to learn from. Through empirical testing, we find that a training SNR of 0 dB for the encoder and -2 dB for the decoder works best. Intuitively, this can be interpreted as learning a good decoder being much harder than learning a good encoder, as the decoder has to work with noisy data, whereas the encoder always takes clean input.

Continuing on this theme, we also observe that training the encoder for fewer iterations is sufficient compared to the decoder. In our experiments, we train the encoder 10x fewer steps than the decoder. 

\textbf{High SNR performance.} Obtaining gains over classical codes in the high SNR regime is challenging - this stems from several factors :
Firstly, the error events become exceedingly sparse in the high SNR regime, leading to noisy gradient estimates and unstable training. Secondly, in the high SNR regime, the error rate is primarily governed by the code’s minimum distance. While DeepPolar exhibits remarkable distance properties (closely resembling the Gaussian codebook - \prettyref{fig:distance}), optimizing the minimum distance is challenging given the exponentially large space of codewords.

Nevertheless, high SNR performance can be boosted by finetuning the trained models at substantially high batch sizes. For instance, we finetune the model using a batch size of 200K (a 10x increase from the training phase), at -1dB using a learning rate of $10^{-5}$.

\subsection{Large batch size}
Having a large batch size during training is desirable for stable training. Larger batch sizes will capture the statistics of the underlying distribution more accurately, improving the accuracy of normalization using the mean and variance of the batch. Moreover, a larger batch size also reduces the noise in the gradients computed. In our experiments, we considered a batch size of up to 20000 and further used gradient accumulation techniques to simulate an even larger batch size.

\subsection{Hyperparameter choices for \dpolar(256,37)}

The hyperparamters used for training \dpolar(256,37,$\ell=16$) are listed in \prettyref{tab:hyperparameters}.

\begin{table}[h]
\centering
\begin{tabular}{lc}
\hline
\textbf{Hyperparameter} & \textbf{Value} \\
\hline
Batch size ($B$) & 20,000 \\
Encoder training SNR & 0 dB \\
Decoder training SNR & -2 dB \\
Total epochs ($T$) & 2,000 \\
Encoder steps per epoch ($T_{enc}$) & 20 \\
Decoder steps per epoch ($T_{dec}$) & 200 \\
Encoder learning rate ($l_{enc}$) & $10^{-4}$ \\
Decoder learning rate ($l_{dec}$) & $10^{-4}$ \\
Optimizer & Adam \\
\hline
\end{tabular}
\caption{Hyperparameters used in model training for \dpolar(256,37,$\ell=16$)}
\label{tab:hyperparameters}
\end{table}

\subsection{Architecture for \dpolar(256,37)}\label{app:arch}
\subsubsection{Encoder}
The encoder is a collection of kernels of size $\ell=16$, each of which is modeled by a neural network $g$. The encoder kernel $g$ is responsible for encoding $\ell$ inputs. The architecture for $g$ is given as follows:


\begin{figure}[ht]

\begin{center}

\includegraphics[width=0.4\columnwidth]{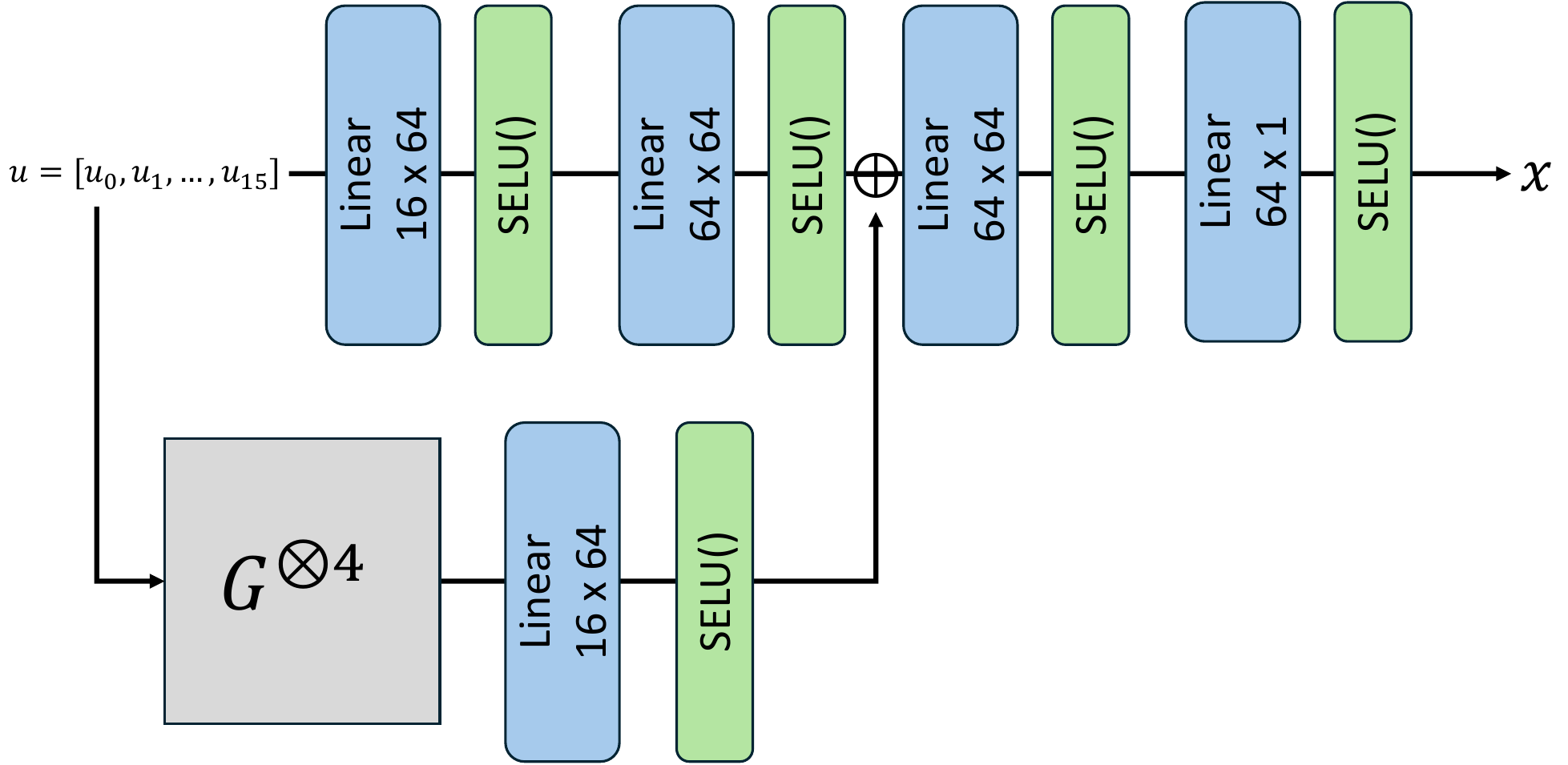}

\caption{Each kernel $g$ at the encoder is parameterized by a fully connected network of 3 hidden layers and a hidden dimension of 64. The features corresponding to the Plotkin transform of size $\ell$ is provided via a skip connection} 
\label{fig:enc_arch}
\end{center}
\end{figure}

A crucial design choice is to make the polar-encoded features available to the encoding network via a skip connection. \dpolar codes rely on the encoding and decoding structures of Polar codes, and the features corresponding to polar codes are indeed informative. Further, since it is harder for NNs to learn multiplicative policies, this side information proves to be useful in the learning process. 

However, it is noteworthy that the learned \dpolar codes, as well as \dpolar-binary codes do not resemble polar codes, both in the mapping as well as the code distance spectrum.



\subsubsection{Decoder}
The encoder is a collection of kernels of size $\ell=16$, each of which is modeled by a neural network $f$. The decoder kernel $f$ contains $\ell=16$ sub-networks $f_i$, each responsible for decoding the kernel's $i^{th}$ position. The architecture for $f_i$ is given as follows:

\begin{figure}[ht]

\begin{center}
\includegraphics[width=0.4\columnwidth]{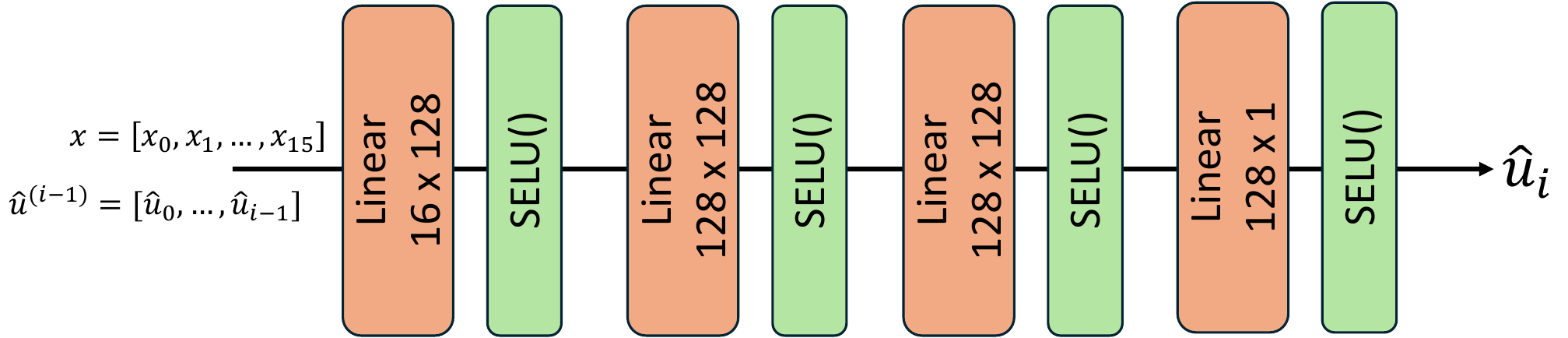}

\caption{Each kernel $f_i$ at the decoder is parameterized by a fully connected network of 3 hidden layers and a hidden dimension of 128. } 
\label{fig:dec_arch}
\end{center}
\end{figure}

\section{Additional results}\label{app:additional}
\subsection{Block Error Rate }\label{app:bler}
\begin{figure*}[ht]
\centerline{

\subfloat[$n=256,k=64$]
{
 \includegraphics[width=0.5\linewidth]{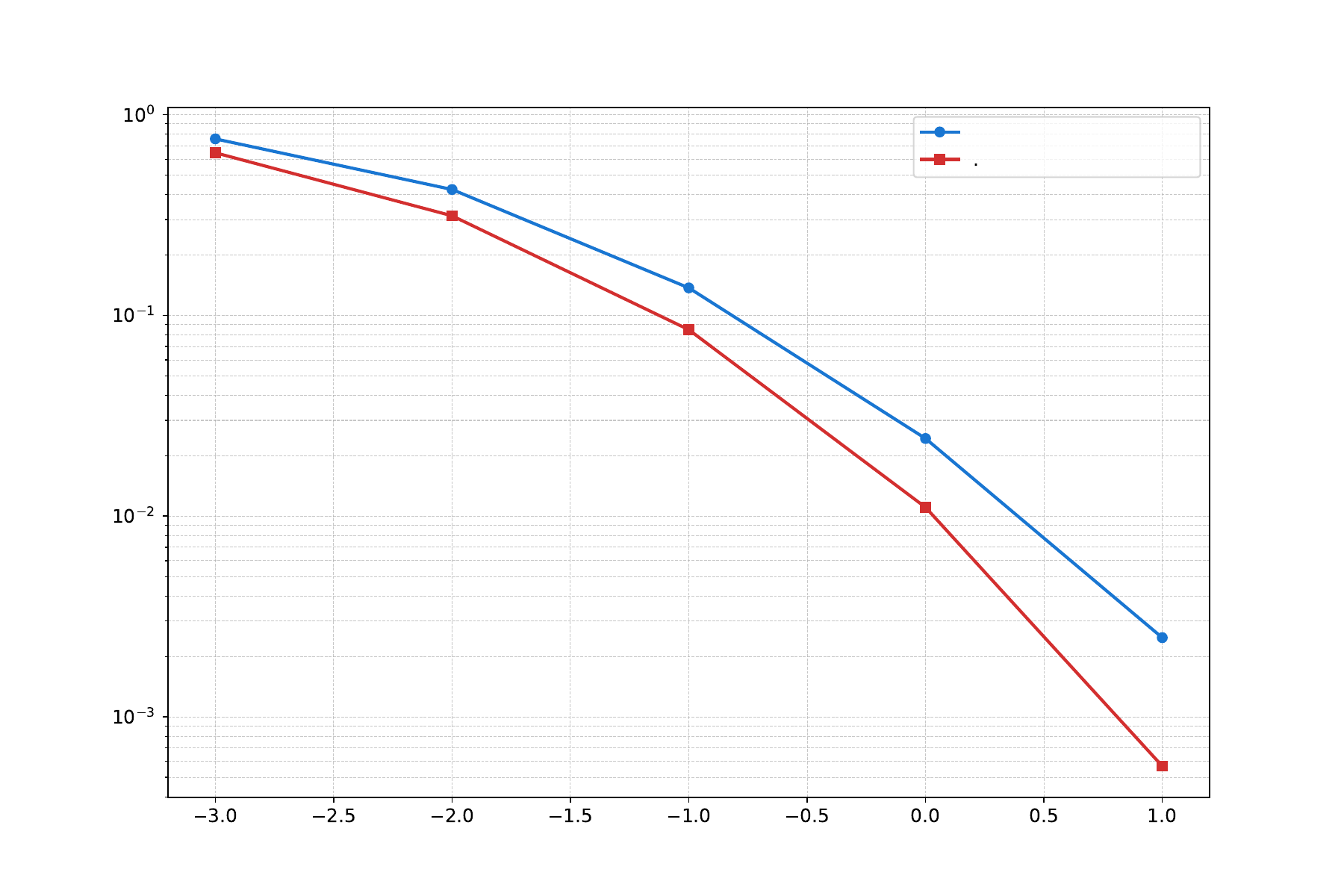}
    \put(-67.5,137.5){\fontsize{3}{5}\selectfont \dpolar }
    \put(-67.5,132.5){\fontsize{3}{5}\selectfont Polar}
  \put(-140,0){\footnotesize SNR (dB)}
\put(-235,70){\rotatebox[origin=t]{90}{\footnotesize Block Error Rate}}
  \label{fig:bler_25664}
}
\hfill
\subfloat[$n=256,k=28$]
{
 \includegraphics[width=0.5\linewidth]{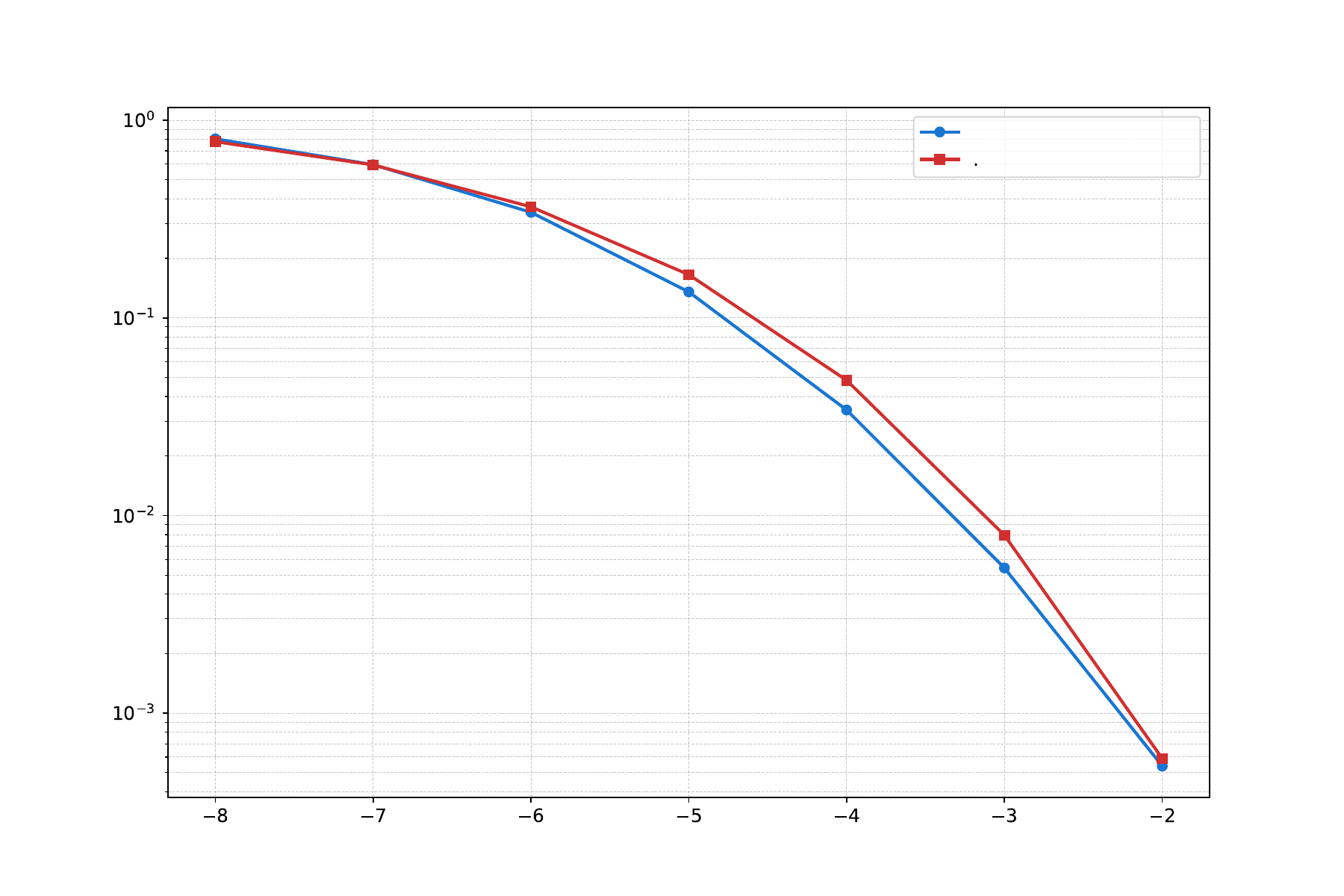}
    \put(-67.5,137.5){\fontsize{3}{5}\selectfont \dpolar}
    \put(-67.5,132.5){\fontsize{3}{5}\selectfont Polar}
  \put(-140,0){\footnotesize SNR (dB)}
\put(-235,70){\rotatebox[origin=t]{90}{\footnotesize Block Error Rate}}
  \label{fig:bler_25628}
}
}
\caption{(a) \dpolar achieves sub-optimal BLER compared to Polar codes: this is an artifact of the BCE training objective, which is a surrogate for the BER}
\label{fig:additional_bler}

\end{figure*}

\begin{figure*}[ht]
\centering{
 \includegraphics[width=0.5\linewidth]{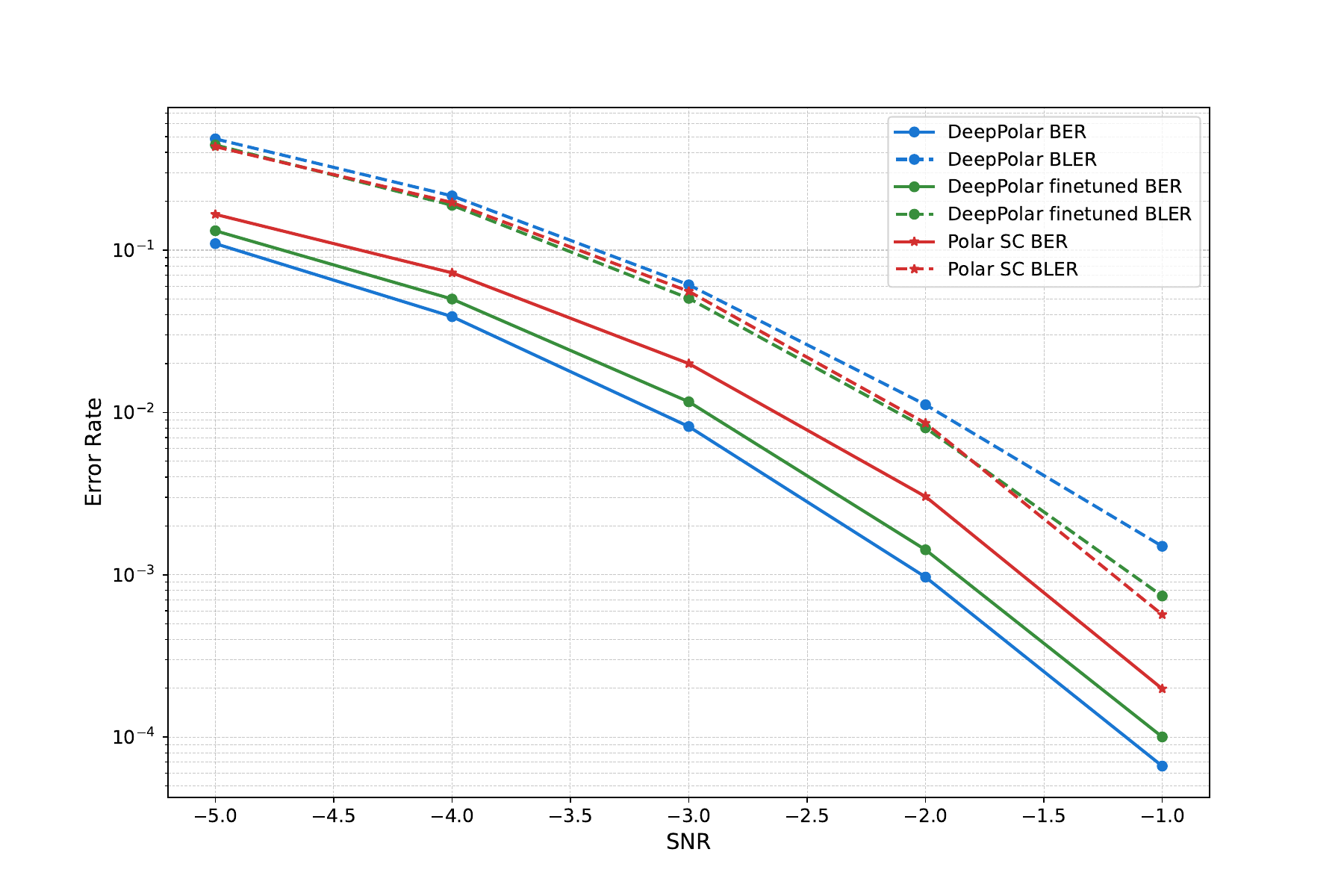}
  \put(-140,0){\footnotesize SNR (dB)}
\put(-235,70){\rotatebox[origin=t]{90}{\footnotesize Error Rate}}
\caption{Finetuning a trained \dpolar model on a BLER surrogate objective trades off BER gains for BLER.}
\label{fig:bler_improvement}
}
\end{figure*}

The figures of merit for any channel code are the Bit Error Rate (BER), and Block Error Rate (BLER). Since these metrics are non-differentiable, we need to find a differentiable surrogate. While the binary cross-entropy loss (used in the paper) acts as a stable surrogate loss function to optimize the BER, identifying such a surrogate loss for the BLER is an unsolved problem. (Often, optimizing for BER results in good BLER (eg - \prettyref{fig:bler_25628}). However, this is not guaranteed).

As preliminary work, we consider several potential loss functions  to optimize the BLER, including $L_1 = 1 - \Pi_{i=1}^{n} \sigma (u_k L_k)$, where $L_k$ is the logit corresponding to the kth bit, and $L_2 = \text{LogSumExp}(x_{1}^{n})$, where $x_k = \text{BCE}_{k}$.

However, both these losses lead to unstable training. For example, the product loss $L_1$ tends to show vanishing gradients. Nevertheless, fine-tuning a model initially trained on BCE loss to optimize the BER, on BLER loss $L_1$, yields improved BLER performance, demonstrated in \prettyref{fig:bler_improvement} This strategy, however, introduces a tradeoff in BER gains - since the BER and BLER objectives are not fully aligned. With the right training objectives, neural codes provide flexibility in optimizing the figures of merit that we are interested in.

\subsection{Model capacity ablation}

\begin{table}[h]
    \centering
    \begin{tabular}{cccccc}
        \hline
        \textbf{Enc hidden size} & \textbf{\# Parameters in Enc} & \textbf{Dec hidden size} & \textbf{\# Parameters in Dec} & \textbf{BER at SNR -1.0 dB} \\
        \hline
        8 & 5K & 8 & 16K & 9e-3 \\
        16 & 12K & 16 & 44K & 7.4e-4 \\
        16 & 12K & 32 & 133K & 2.1e-4 \\
        32 & 33K & 32 & 133K & 8.3e-5 \\
        64 & 100K & 128 & 1.6M & 5.9e-5 \\
        \hline
    \end{tabular}
    \caption{Performance comparison as a function of the number of parameters in the encoder and decoder.}
    \label{tab:model_capacity}
\end{table}

\begin{figure*}[ht]
\centerline{
\subfloat[]
{  \hspace*{-0.25in}
 \includegraphics[width=0.5\linewidth]{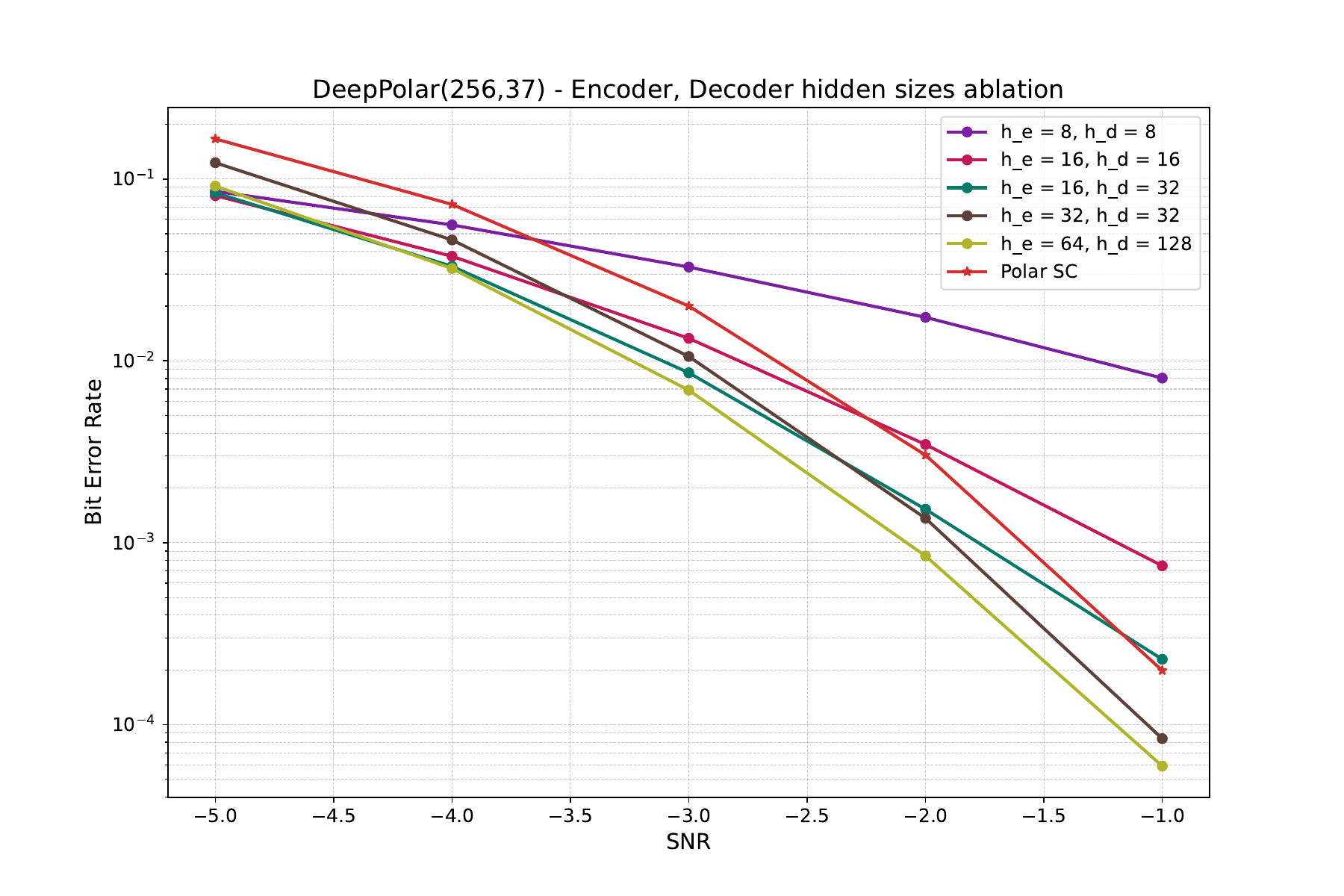} 
    \put(-140,0){\footnotesize SNR (dB)}
\put(-235,70){\rotatebox[origin=t]{90}{\footnotesize Bit Error Rate}}\label{fig:ber_model_capacity}
 }
\hfill
\subfloat[]
{
 \includegraphics[width=0.5\linewidth]{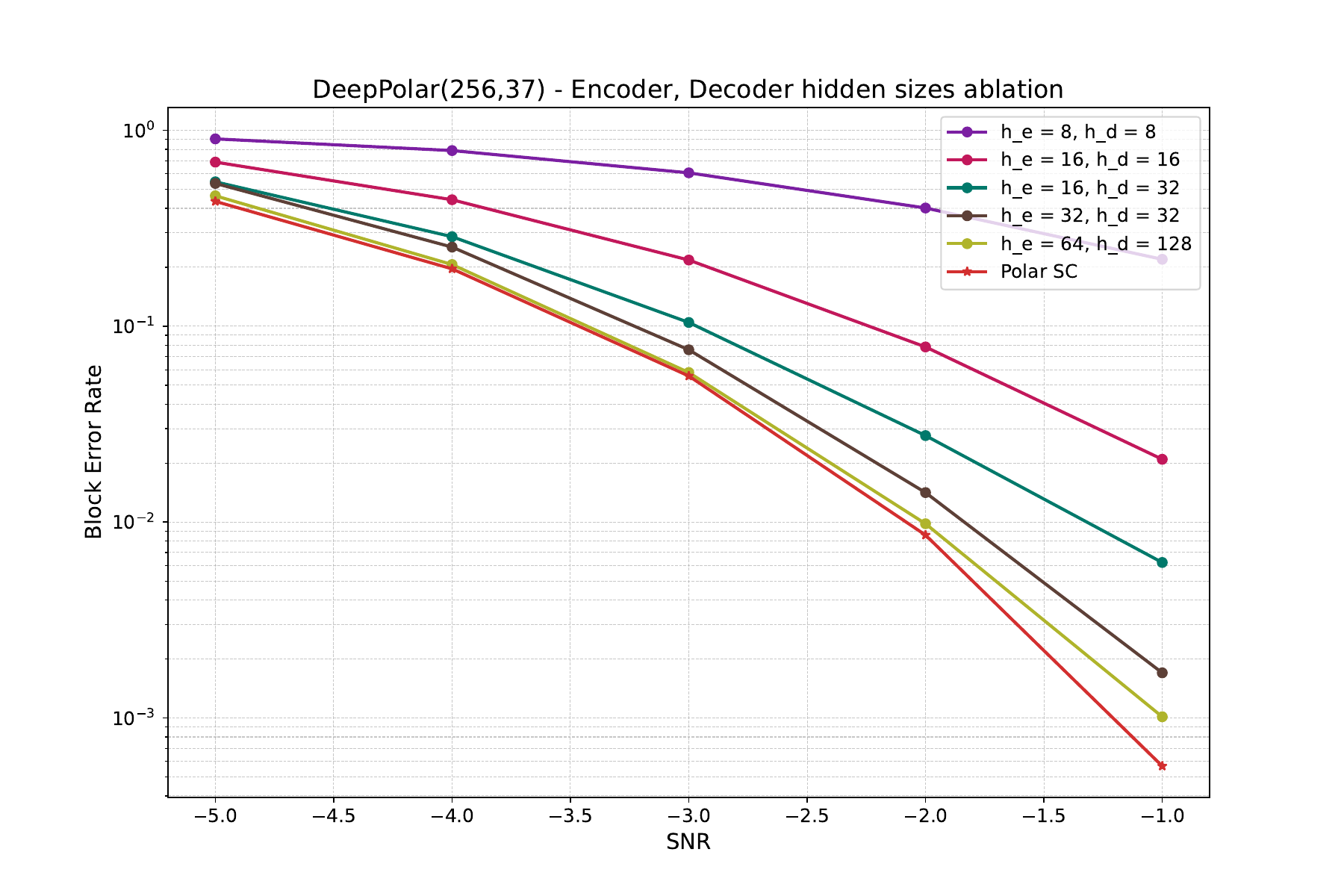}
    \put(-140,0){\footnotesize SNR (dB)}
\put(-235,70){\rotatebox[origin=t]{90}{\footnotesize Block Error Rate}}\label{fig:bler_model_capacity}
}
}
\caption{Performance as a function of model size with pretraining and curriculum training: \dpolar($256,37,\ell=16$) }
\label{fig:model_capacity}

\end{figure*}

\dpolar-SC consists of a total of 100K+1.6M parameters; each kernel is an NN of depth 3, with hidden sizes of 64 and 128, respectively, at the encoder and decoder. We use the notation $(e=64,d=128)$ to represent this model. Generally, overparameterized models are easier to train. Since model size is an important practical consideration, we study the effect of model size on reliability. We consider the encoder and decoder networks with kernels with hidden sizes $e$ and $d$ respectively with depth $3$. We train the networks via kernel pretraining and curriculum learning, as outlined in \prettyref{sec:curriculum}. As shown in \prettyref{tab:model_capacity}, \prettyref{fig:model_capacity}, there is a noticeable degradation in performance when small models, i.e., $(e=8,d=8)$, $(e=16,e=16)$, are used to parameterize each kernel. However, $(e=16,e=32)$, and specifically $(e=32,h=32)$ with 160K parameters (10x reduction), outperforms SC in terms of BER and approaches the performance of $(e=64,d=128)$. These observations can be attributed to the dimension of the input, which is the kernel size 16.



\subsection{Fully connected neural code - lacks generalization.} \label{app:fcnn}
\begin{figure*}[h]
\centering{
 \includegraphics[width=0.5\linewidth]{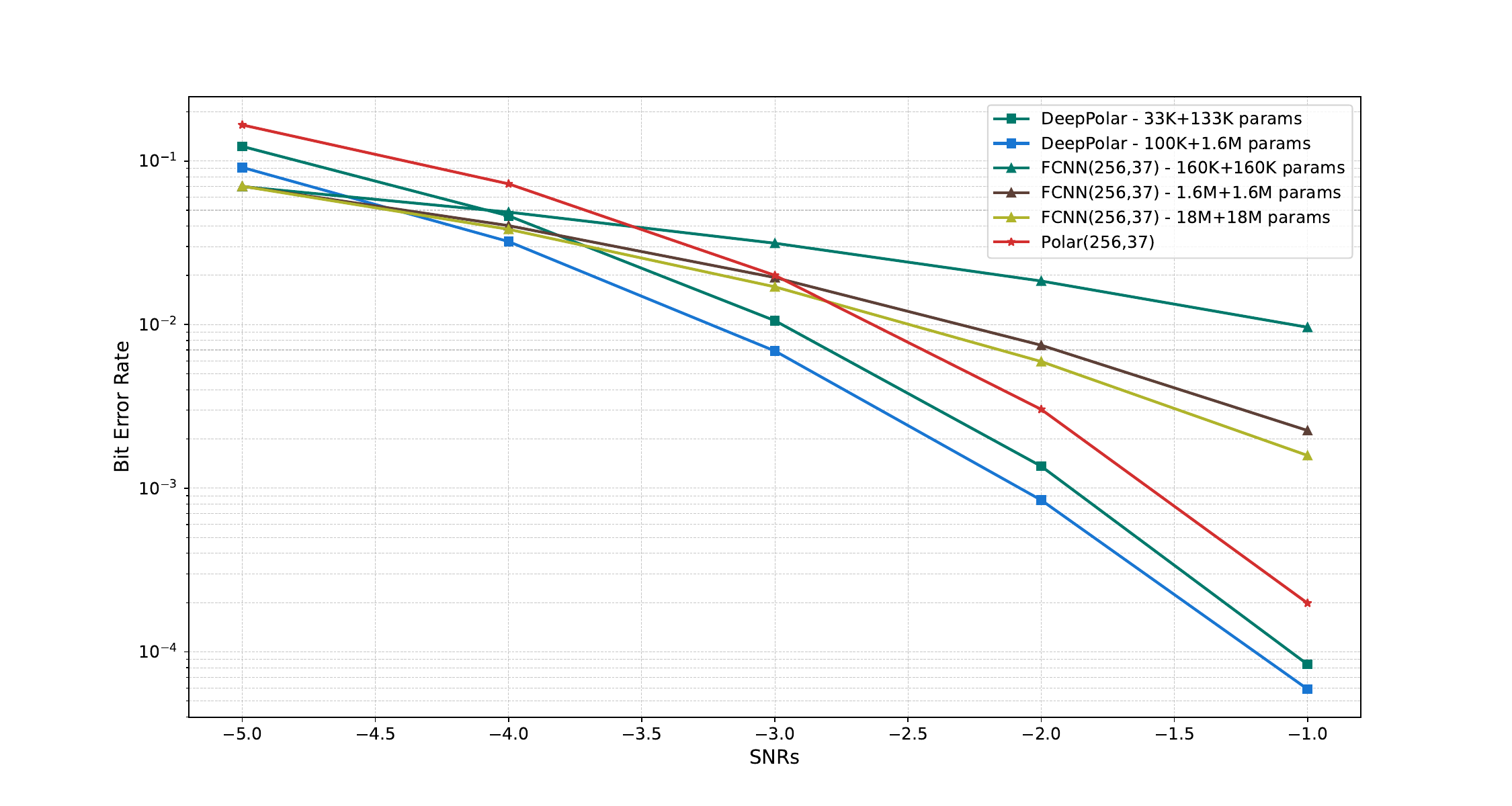}
\caption{Codes learned via Fully connected NNs fail to generalize to unseen messages due to the absence of strong inductive biases.}
\label{fig:fcnn}
}
\end{figure*}

While it might seem intuitive that non-linear codes would give better results compared to linear codes, the construction of good non-linear codes is highly non-trivial. To illustrate the difficulty, we train a dense, fully connected encoder-decoder pair using the same training methodology as \dpolar (\prettyref{sec:experimental}). As shown in \prettyref{fig:fcnn}, despite this network having ten times more parameters than \dpolar, its performance is substantially inferior to the successive cancellation (SC) decoding of polar codes. This is an effect of the model's poor generalization in the absence of good inductive biases, a phenomenon well-documented in existing literature, for example, see \cite{jiang2020learn}. Thus, neural architectures that add redundancy in a structured way are essential for learning a good code.

\end{document}

%% file: commands.tex
\usepackage[utf8]{inputenc} 
\usepackage[T1]{fontenc}    
\usepackage{url}            
\usepackage{booktabs}       
\usepackage{amsfonts}       
\usepackage{nicefrac}       
\usepackage{microtype}      
\usepackage{tabu}
\usepackage{multicol}
\usepackage{soul}
\usepackage{bbm}
\usepackage{lipsum}
\usepackage{kantlipsum}

\usepackage{cancel}

\usepackage{amsmath,amssymb,amsfonts,amsthm, dsfont, color}
\usepackage{algorithm}
\usepackage{mathtools}
\usepackage{graphicx}
\usepackage{textcomp}
\usepackage{natbib}
\usepackage{xcolor, fancyhdr, xspace}
\usepackage{enumerate}
\usepackage{float}

\usepackage{mathtools}
\usepackage{cuted}

\definecolor{MyGreen1}{RGB}{20,180,40}
\usepackage{multirow, tikz,float}

\usepackage{nicefrac,color,mathrsfs,float}
\usepackage{multirow,caption,tikz}
\usetikzlibrary{shapes.misc, positioning}
\usetikzlibrary{decorations.pathreplacing}

\tikzset{
  block/.style    = {draw, thick, rectangle, minimum width = 2em},
sblock/.style      = {draw, thick, rectangle, minimum height = 2em,
minimum width = 2em}, 
}

\newcommand{\polar}{\text{Polar}}
\newcommand{\dpolar}{\textsc{DeepPolar}\xspace}

\usepackage{comment}

\usepackage{thmtools}

\usepackage{prettyref,xspace}
\usepackage{tikz}

\newrefformat{cond}{Condition~\ref{#1}}
\newrefformat{eq}{Eq.~\ref{#1}}
\newrefformat{thm}{Theorem~\ref{#1}}
\newrefformat{th}{Theorem~\ref{#1}}
\newrefformat{chap}{Chapter~\ref{#1}}
\newrefformat{sec}{Sec.~\ref{#1}}
\newrefformat{algo}{Algorithm~\ref{#1}}
\newrefformat{fig}{Fig.~\ref{#1}}
\newrefformat{tab}{Table~\ref{#1}}
\newrefformat{rmk}{Remark~\ref{#1}}
\newrefformat{clm}{Claim~\ref{#1}}
\newrefformat{def}{Definition~\ref{#1}}
\newrefformat{cor}{Corollary~\ref{#1}}
\newrefformat{lmm}{Lemma~\ref{#1}}
\newrefformat{prop}{Proposition~\ref{#1}}
\newrefformat{pr}{Proposition~\ref{#1}}
\newrefformat{app}{App.~\ref{#1}}
\newrefformat{prob}{Problem~\ref{#1}}
\newrefformat{ques}{Question~\ref{#1}}
\newrefformat{notee}{Note~\ref{#1}}
\newrefformat{assump}{Assumption~\ref{#1}}
\newrefformat{issuee}{Issue ~\ref{#1}}
\newrefformat{fix}{Fix ~\ref{#1}}

\newcommand{\reals}{\mathbb{R}}

\newcommand{\pprob}[1]{\mathbb{P}[#1]}

\newcommand{\vect}[1]{\boldsymbol{#1}}

\newcommand{\bma}{\vect{m}}

\newcommand{\bu}{\vect{u}}

\newcommand{\bx}{\vect{x}}
\newcommand{\by}{\vect{y}}

\newcommand{\binary}{\{0,1\}}

\newcommand{\define}{\triangleq}

\newcommand{\ie}{i.e.\xspace}

\mathchardef\mhyphen="2D

\definecolor{MyGreen1}{RGB}{20,180,40}

\usepackage{siunitx}

